\documentclass[twocolumn,showkeys,showpacs,preprintnumbers,amsmath,amssymb,floatfix,prd]{revtex4}

\usepackage{graphicx}
\usepackage{dcolumn}
\usepackage{bm}

\newcommand{\beq}{\begin{equation}} 
\newcommand{\eeq}{\end{equation}}
\newcommand{\bey}{\begin{eqnarray}}
\newcommand{\eey}{\end{eqnarray}}

\begin{document}

\preprint{}

\title{\bf  Black holes/naked singularities in four-dimensional non-static space-time and the energy-momentum distributions}

\author{\bf Faizuddin Ahmed}
 \email{ faizuddinahmed15@gmail.com}
\affiliation {\bf Ajmal College of Arts and Science, Dhubri-783324, Assam, India}

\author{\bf Farook Rahaman}
\email{rahamank@associates.iucaa.in}

\author{\bf Susmita Sarkar}
 \email{ susmita.mathju@gmail.com}
\affiliation {\bf Department of Mathematics, Jadavpur University, Kolkata-700032, India}

\date{\today}

\begin{abstract}

In this article, we discuss four dimensional non-static space-times in the background of de-Sitter and anti-de Sitter spaces with the matter-energy sources a stiff fluid, anisotropic fluid, and an electromagnetic field. Under various parameter conditions the solutions may represent models of naked singularity and/or black holes. Finally, the energy-momentum distributions using the complexes of Landau-Lifshitz, Einstein, Papapetrou, and M{\o}ller prescriptions, were evaluated.

\end{abstract}


\maketitle

\section{Introduction}

The formation of space-time singularities is a common phenomena in General Relativity. However, it is widely believed that space-time singularities do not exist in Nature, they represent a limitation of the classical theory. The earliest idea, in mid sixties, by Sakharov \cite{Sakharov} and Gliner \cite{Gliner} suggested that singularities could be avoided by matter, {\it i. e.,} with a de-Sitter core, with equation of state $p=-\rho$. The fundamental features of a black hole is the topology. In 4D asymptotically flat stationary space-times, Hawking first showed that a black hole has necessarily a $S^{2}$ topology, provided that the dominant energy condition (DEC) holds \cite{Hawk}. Later, it was realized that Hawking's theorem can be improved in various aspects (see \cite{Friedmann,Friedmann2,Galloway,Cai}). However, once the energy condition is relaxed, a black hole can have quite different topologies. Such examples can occur even in four-dimensional space-times where the cosmological constant is negative \cite{Huang,Lemos,Cai2,Mann,Brill, Vanzo,Cai3,Klemm,Klemm2}. It is worth to mentioning that the formation of topological black holes from gravitational
collapse in four-dimensional space-times were studied in \cite{Smith,RG,JLemos}.

The curvature singularities are the singularities where the Riemann tensor components or the zeroth and higher order derivative of scalar curvature diverge, that is, infinite. For black hole solution like Schwarzschild space-time, the curvature singularity occurs at $r=0$ is covered by an event horizon exist at $r=2\,M$ where, $M$ represents mass parameter. On the other hand, in some other solutions the curvature singularity is not covered by any event horizon. Generally the curvature singularities of nonvacuum (or matter-filled) as well as the vacuum space-times are recognized from the divergence of the energy-density and/or the scalar curvature, such as the zeroth order scalar curvature $R_{\mu\nu\rho\sigma}\,R^{\mu\nu\rho\sigma}$ (called the Kretschmann scalar) and $R_{\mu\nu\rho\sigma}\,R^{\rho\sigma\lambda\tau}\,R^{\mu\nu}_{\lambda\tau}$. In addition, by analysing the outgoing radial null geodesics of a space-time containing space-time singularity, one can determine whether the curvature singularity is naked or covered by an event horizon (see Refs. \cite{Jos1,Jos2} for detail discussion). For the strength of curvature singularities, two conditions are used : one is the {\it strong curvature condition} (SCC) given by Tipler \cite{Tipler}, and the other is the {\it limiting focusing condition} (LFC) given by Krolak \cite{Kro}. Meanwhile the curvature singularities are basically of three types : first one being a space-like singularity ({\it e. g.} Schwarzschild singularity), second one the time-like singularity, and third one is the null singularity. In time-like singularity, two possibilities arise : $(i)$ there is an event horizon around a time-like singularity ({\it e. g.} R-N black holes). Here an observer cannot see a time-like singularity from an outside region of the space-time, that is, the singularity is covered by an event horizon ; $(ii)$ there is no event horizon around a time-like singularity which we called the naked singularity (NS), and it would observable for far away or distant observers. Therefore a naked singularity may present opportunities to observe the physical effects near the very dense regions that formed in the very final stages of a gravitational collapse. But in a black hole scenario, such regions are necessarily hidden within the event horizon. There are many example of gravitational collapse models which formed a naked singularity known. The earliest model is the Lemaitre-Tolman-Bondi (LTB) \cite{Lei,Tol,Bondi} solutions, a spherically symmetric inhomogeneous collapse of dust fluid that admits both naked and covered singularity. Other spherically symmetric gravitational collapse space-times which formed a naked singularity would be \cite{pap,vai,Christ,Des,Gov,Brave,Clarke, Jos3,Jos4,Glass,Rocha,Herra,Kras}. To counter the occurence of naked singularity in a solution of the Einstein's field equations, Penrose proposed a Cosmic Censorship Conjecture (CCC) \cite{Pen1,Pen2,Pen3}. However, the general and/or detail proofs of this Conjecture has not yet been given. On the contrary, there is no mathematical details yet known which forbid the appearence of naked singularity in a solution of the field equations.

In addition to the spherical symmetric system, non-spherical gravitational collapse space-times with naked singulairty are also of particular interest in general relativity. The Gravitational collapsing models in cylindrical symmetry system which formed naked singularities has been discussed by Thorne \cite{Thor} and Hayaward \cite{Hay}. After that several authors investigated the gravitational collapse solutions in cylindrical symmetry system (for examples, \cite{Apo,Eche,Gutt,Nakao,Nakao1,Gonc,Nolan,Peri,Wang,Faiz5,Faiz6}). Some other examples of non-spherical gravitational collapse solutions with naked singularities would be \cite{Chi,Piran,Th,Morgan,Lete,JMM,Bondi2,Mel,Wang2,Faiz,Faiz2, Faiz3,Faiz4,Faiz7}. Attempts have been made to distinguish between black holes and naked singularities from astrophysical data through gravitational lensing. In this direction, most significant works have been done by Virbhadra and his collaborators, for examples, in the Janis-Newman-Winicour space-time \cite{KS,KS2} (see also \cite{Chou,Zhou}), its rotating generalization \cite{Yaz}, and the work in \cite{Sahu}. There have been other attempts to explore the physical applications and implications of naked singularities (see \cite{Mala,Mala2} and references therein). The recent proposal to consider naked singularities as possible particle accelerators were in \cite{Patil,Patil2}. In addition, some other valuable work done by several authors would be in \cite{Wer,Bambi,Bambi2,Hioki}. Other relativist had shown that naked singularities and black holes can be differentiated by the properties of accretion disks that accumulate around them (see \cite{Chou,Nara} and references therein).

The Einstein's field equations with cosmological constant are given by
\begin{equation}
G^{\mu}+\Lambda\,g_{\mu\nu}=\kappa\,T^{\mu\nu},\quad \mu,\nu=0,1,2,3,
\label{Ein}
\end{equation}
where $\kappa=8\,\pi$ and $T^{\mu\nu}$ is the stress-energy tensor. For our present work, we have chosen the following stress-energy tensors:

(i) Perfect fluid :
\begin{equation}
T^{0}_{0}=-\rho,\quad T^{1}_{1}=T^{2}_{2}=T^{3}_{3}=p,
\label{perfect}
\end{equation}
where $\rho$ is the energy-density and $p$ is the isotropic pressure.

(ii) Anisotropic fluid :
\begin{equation}
T^{0}_{0}=-\rho,\quad T^{1}_{1}=p_{x},\quad T^{2}_{2}=p_{y},\quad T^{3}_{3}=p_{z},
\label{anisotropic}
\end{equation}
where $\rho$ is the energy-density and $p_{i}$, where $i=x,y,z$ the pressures.

(iii) Electromagentic field :
\begin{equation}
T_{\mu\nu}=\frac{2}{\kappa}\,(g^{\alpha\beta}\,F_{\alpha\mu}\,F_{\beta\nu}-\frac{1}{4}\,g_{\mu\nu}\,F_{\alpha\beta}\,F^{\alpha\beta}).
\label{em}
\end{equation}
where $F_{\mu\nu}=\partial_{\mu} A_{\nu}-\partial_{\nu} A_{\mu}$ is the electromagentic field tensor satisfying the following conditions
\begin{equation}
F^{\mu\nu}_{\,\,\,;\nu}=0,\quad F_{[\mu\nu;\alpha]}=0,
\label{}
\end{equation}
for the source-free regions, where $A_{\mu}$ is the four-vector potential.

The different energy conditions \cite{Hawking,Wald,Steph} are as follows:
\begin{eqnarray}
&&WEC\quad\quad:\quad\quad \rho>0,\quad \rho+p_{i}\geq 0,\nonumber\\
&&SEC\quad\quad:\quad\quad \rho+p_{i}\geq 0,\quad \rho+\sum_{i} p_{i}\geq 0,\nonumber\\
&&DEC\quad\quad:\quad\quad \rho\geq |p_{i}|,\quad i=1,2,3.
\label{energy-conditions}
\end{eqnarray}

In the present work, taking into account the energy conditions, we attempt to construct four-dimensional non-static space-time which may represent models of naked singularities and/or black holes solution under various parameter conditions. Finally, we evaluate the energy-momentum distributions using the known energy-momentum complexes.

\section{Analysis of the space-time}

The four dimensional non-static metric \cite{Wu} is given by
\begin{eqnarray}
ds^2&=&g_{\mu\nu}\,dx^{\mu}\,dx^{\nu}\nonumber\\
&=&-f(x)\,dt^2+\frac{1}{f(x)}\,dx^2+x^2\,[A(t)\,dy^2+\frac{1}{A(t)}\,dz^2],\quad
\label{1}
\end{eqnarray}
where
\begin{equation}
f(x)=(\frac{2\,\beta}{x}+\frac{\alpha^2}{x^2}-\frac{\Lambda}{3}\,x^2+\delta\,x),\quad A(t)=e^{-2\,\gamma\,t}.
\label{2}
\end{equation}
Here $\beta$ represents mass parameter, $\alpha$ represents charge, $\Lambda$ is the cosmological constant, $\delta$ and $\gamma$ are integer. The metric functions with its inverse are
\begin{eqnarray}
g_{00}&=&-f(x)=\frac{1}{g^{00}},\quad g_{11}=\frac{1}{f(x)}=\frac{1}{g^{11}},\nonumber\\
g_{22}&=&x^2\,e^{-2\,\gamma\,t}=\frac{1}{g^{22}},\quad g_{33}=x^2\,e^{2\,\gamma\,t}=\frac{1}{g^{33}}.
\label{metric-functions}
\end{eqnarray}
The metric is Lorentzian with signature $(-,+,+,+)$ and the determinant of the corresponding metric tensor $g_{\mu\nu}$ is
\begin{equation}
det\;g=-x^4.
\label{3}
\end{equation}
Thus the metric is regular everywhere except at $x=0$. The nonzero components of the Einstein tensor are
\begin{eqnarray}
G^{0}_{0}&=&-\frac{\alpha^2}{x^4}+\frac{2\,\delta}{x}-\Lambda+\frac{\gamma^2}{f(x)},\nonumber\\
G^{1}_{1}&=&-\frac{\alpha^2}{x^4}+\frac{2\,\delta}{x}-\Lambda-\frac{\gamma^2}{f(x)},\nonumber\\
G^{2}_{2}=G^{3}_{3}&=&\frac{\alpha^2}{x^4}+\frac{\delta}{x}-\Lambda-\frac{\gamma^2}{f(x)},
\label{4}
\end{eqnarray}

The time-like four velocity vector when $t$ is time-like and $x$ is space-like defined by
\begin{equation}
u^{\mu}={f(x)}^{-1/2}\,\delta^{\mu}_{t},\quad u^{\mu}\,u_{\mu}=-1,
\label{5}
\end{equation}
such that the four-acceleration vector is
\begin{equation}
a^{\mu}=u^{\mu\,;\,\nu}\,u_{\nu}=k(x)\,\delta^{\mu}_{x},
\label{6}
\end{equation}
where the quantity $k(x)=\frac{1}{2}\,f'(x)$ is called the surface gravity for a static observer. Here prime deonte derivative w. r. t. $x$.

We study the following cases of the above non-static spacetime.

\begin{center}
{\bf Case 1} : $\beta\neq 0$, $\alpha=0$, $\Lambda=0$, $\delta=0$, $\gamma\neq 0$.
\end{center}

The curvature invariant given by
\begin{equation}
R^{\mu\nu\rho\sigma}\,R_{\mu\nu\rho\sigma}=\frac{48\,\beta^2}{x^6}-\frac{20\,\gamma^2}{x^2}+\frac{3\,\gamma^4\,x^2}{\beta^2},
\label{7}
\end{equation}
diverges at $x=0$ without covered by an event horizon and thus a naked singularity is formed.

Therefore from the field equations using (\ref{4}) and (\ref{perfect}) we get
\begin{equation}
\kappa\,\rho=\kappa\,p=-\frac{\gamma^2\,x}{2\,\beta},
\label{8}
\end{equation}
a stiff fluid, where $\kappa=8\,\pi$. If one takes $\gamma=0$, this corresponds to the Taub vacuum solution \cite{Taub} and the properties of which are well-known.

For $\gamma \neq 0$, there are four cases arise as follow :

\vspace{0.1cm}

(i) $\beta>0$, $x>0$

Here $t$ is time-like and $x$ is space-like coordinate and the energy density of stiff fluid violate the weak energy conditions.

\vspace{0.1cm}

(ii) $\beta>0$, $x<0$

In that case, $t$ is space-like and $x$ is time-like coordinate. So, the energy density of stiff fluid satisfy the energy conditions.

\vspace{0.1cm}

(iii) $\beta<0$, $x<0$

In that case, $t$ is time-like and $x$ is space-like coordinate. Here the energy density violate the energy conditions.

\vspace{0.1cm}

(iv) $\beta<0$, $x>0$

In that case, $t$ is space-like and $x$ is time-like coordinate. So the energy density satisfy the energy conditions.

\begin{center}
{\bf Case 2} : $\beta=0$, $\alpha\neq 0$, $\Lambda=0$, $\delta=0$, $\gamma\neq 0$.
\end{center}

The curvature invariant in that case is given by
\begin{equation}
R^{\mu\nu\rho\sigma}\,R_{\mu\nu\rho\sigma}=\frac{56\,\alpha^4}{x^8}-\frac{40\,\gamma^2}{x^2}+\frac{12\,\gamma^4\,x^4}{\alpha^4},
\label{9}
\end{equation}
which diverges at $x=0$ not covered by an event horizon and thus a naked singularity is formed.

From the field equations using (\ref{4}) and (\ref{anisotropic}) we get
\begin{equation}
\kappa\,\rho=\frac{\alpha^2}{x^4}-\frac{\gamma^2\,x^2}{\alpha^2},\quad \kappa\,p_{x}=-\frac{\alpha^2}{x^4}-\frac{\gamma^2\,x^2}{\alpha^2},\quad p_{y}=p_{z}=\rho.
\label{10}
\end{equation}
Here $t$ is time-like and $x$ is a space-like coordinate. Writing $x_{0}=(\frac{\alpha^2}{\gamma})^{\frac{1}{3}}$, the matter-energy content anisotropic fluid satisfy the following energy conditions:
\begin{eqnarray}
&&WEC\quad : \quad\quad\quad \rho>0,\nonumber\\
&&WEC_{x}\quad :\quad\quad \rho+p_{x}<0,\nonumber\\
&&WEC_{y}\quad :\quad\quad \rho+p_{y}>0,\quad x<x_{0}\nonumber\\
&&WEC_{z}\quad :\quad\quad \rho+p_{z}>0,\quad x<x_{0}\nonumber\\
&&SEC\quad\quad:\quad\quad \rho+\sum_{i} p_{i}>0,\quad 0<(\frac{x}{x_0})^{6}<\frac{1}{2}\nonumber\\
&&SEC\quad\quad:\quad\quad \rho+\sum_{i} p_{i}<0,\quad \frac{1}{2}<(\frac{x}{x_0})^{6}<1\nonumber.
\end{eqnarray}

\vspace{0.2cm}
\begin{center}
{\bf Case 3} : $\beta=0$, $\alpha=0$, $\Lambda\neq 0$, $\delta=0$, $\gamma\neq 0$.
\end{center}

The curvature invariant in that case is given by
\begin{equation}
R^{\mu\nu\rho\sigma}\,R_{\mu\nu\rho\sigma}=\frac{8\,\Lambda^2}{3}+\frac{108\,\gamma^4}{x^4\,\Lambda^2}-\frac{8\,\gamma^2}{x^2},
\label{11}
\end{equation}
which diverges at $x=0$ not covered by an event horizon and thus a naked singularity is formed.

For the field equations using (\ref{4}) and (\ref{perfect}), we get
\begin{equation}
\kappa\,\rho=\kappa\,p=\frac{3\,\gamma^2}{x^2\,\Lambda}, \quad \Lambda<0 \quad \mbox{or} >0.
\label{12}
\end{equation}

There are two possibilities arise :

\vspace{0.1cm}

(i) For $\Lambda>0$, $t$ is space-like and $x$ is time-like coordinate. Therefore the matter-energy sources stiff fluid obeying the energy conditions diverge at $x=0$. Thus the solution represents a stiff fluid in the backgrounds of de Sitter (dS) spaces with a naked singularity.

\vspace{0.1cm}

(ii) For $\Lambda<0$, $t$ is time-like and $x$ is space-like coordinate. In that case, the stiff fluid violates the weak energy condition (WEC). Therefore, the solution represents a stiff fluid in the backgrounds of anti-de Sitter (AdS) space with a naked singularity violate the energy conditions.

Note that if one takes $\gamma=0$, then the space-time represent de-Sitter or anti-de Sitter space, depending on the sign of the cosmological constant $\Lambda$ whose global structure were studied extensively in \cite{Hawking}.

\vspace{0.2cm}
\begin{center}
{\bf Case 4} : $\beta=0$, $\alpha=0$, $\Lambda=0$, $\delta\neq 0$, $\gamma\neq 0$.
\end{center}

In that case, the metric (\ref{1}) reduces to the following form
\begin{equation}
ds^2=-\delta\,x\,dt^2+\frac{1}{\delta\,x}\,dx^2+x^2\,(e^{-2\,\gamma\,t}\,dy^2+e^{2\,\gamma\,t}\,dz^2).
\label{case4}
\end{equation}
Transforming the above metric by $x\rightarrow e^{\delta\,{\bar x}}$, we get
\begin{equation}
ds^2=\delta\,e^{\delta\,{\bar x}}\,(-dt^2+d{\bar x}^2)+e^{2\,(\delta\,{\bar x}-\gamma\,t)}\,dy^2+e^{2\,(\delta\,{\bar x}-\gamma\,t)}\,dz^2.
\label{case4-transform}
\end{equation}
The non-zero components of the Einstein tensor are
\begin{eqnarray}
G^{0}_{0}&=&\delta\,e^{-\delta\,{\bar x}}\,(2+\frac{\gamma^2}{\delta^2}),\nonumber\\
G^{1}_{1}&=&\delta\,e^{-\delta\,{\bar x}}\,(2-\frac{\gamma^2}{\delta^2}),\nonumber\\
G^{2}_{2}&=&\delta\,e^{-\delta\,{\bar x}}\,(1-\frac{\gamma^2}{\delta^2})=G^{3}_{3}.
\label{case4-Einstein}
\end{eqnarray}
Therefore the scalar curvature invariant is given by
\begin{equation}
R^{\mu\nu\rho\sigma}\,R_{\mu\nu\rho\sigma}=4\,e^{-2\,\delta\,{\bar x}}\,(2\,\delta^2-\gamma^2+\frac{3\,\gamma^4}{\delta^2}).
\label{13}
\end{equation}
Consider the stress-energy tensor by (\ref{anisotropic}), from the field equations using (\ref{case4-Einstein}) we get
\begin{eqnarray}
-\kappa\,\rho&=&\delta\,e^{-\delta\,{\bar x}}\,(2+\frac{\gamma^2}{\delta^2}),\nonumber\\
\kappa\,p_{x}&=&\delta\,e^{-\delta\,{\bar x}}\,(2-\frac{\gamma^2}{\delta^2}),\nonumber\\
\kappa\,p_{y}&=&\delta\,e^{-\delta\,{\bar x}}\,(1-\frac{\gamma^2}{\delta^2})=\kappa\,p_{z}.
\label{14}
\end{eqnarray}
From above we see that the scalar curvature invariant (\ref{13}) and the physical parameters (\ref{14}) diverge at ${\bar x}\rightarrow -\infty$ ($x\rightarrow 0$) and vanish rapidly at ${\bar x}\rightarrow \infty$ ($x\rightarrow \infty$). Thus the space-time formed a naked singularity not covered by an event horizon.

In the following we discuss two possibilities:

({\bf A}) If we set $\gamma=\delta$, then from eqn. (\ref{14}) we get
\begin{equation}
-\kappa\,\rho=3\,\delta\,e^{-\delta\,{\bar x}},\quad \kappa\,p_{x}=\delta\,e^{-\delta\,{\bar x}},\quad p_{y}=0=p_{z}.
\label{15}
\end{equation}
The scalar curvature invariant given by
\begin{equation}
R^{\mu\nu\rho\sigma}\,R_{\mu\nu\rho\sigma}=8\,e^{-2\,\delta\,{\bar x}}
\label{16}
\end{equation}
diverge at ${\bar x}\rightarrow -\infty$ ($x\rightarrow 0$) and vanish rapidly at ${\bar x}\rightarrow \infty$ ($x\rightarrow \infty$). The matter-energy content satisfy the different energy conditions for $\delta<0$ and violate for $\delta>0$.

({\bf B}) if we set $\gamma=0$, then the space-time (\ref{1}) represents a conformally flat static solution with anisotropic fluid as the stress-energy tensor. From (\ref{case4-transform}) we get
\begin{equation}
ds^2=\delta\,e^{\delta\,{\bar x}}\,(-dt^2+d{\bar x}^2)+e^{2\,\delta\,{\bar x}}\,(dy^2+dz^2).
\label{m1}
\end{equation}
The curvature invariant given by
\begin{equation}
R^{\mu\nu\rho\sigma}\,R_{\mu\nu\rho\sigma}=8\,\delta^2\,e^{-2\,\delta\,{\bar x}},
\label{m2}
\end{equation}
which diverge at ${\bar x}\rightarrow -\infty$ ($x\rightarrow 0$) and vanish rapidly at ${\bar x}\rightarrow +\infty$ ($x\rightarrow +\infty$).

The matter-energy content anisotropic fluid from (\ref{14})
\begin{eqnarray}
-\kappa\,\rho&=&2\,\delta\,e^{-\delta\,{\bar x}},\nonumber\\
\kappa\,p_{x}&=&2\,\delta\,e^{-\delta\,{\bar x}},\nonumber\\
\kappa\,p_{y}&=&\delta\,e^{-\delta\,{\bar x}}=\kappa\,p_{z}.
\label{m3}
\end{eqnarray}
The matter-energy content anisotropic fluid satisfy the following energy conditions :
\begin{eqnarray}
&&WEC\quad\quad : \quad\quad\quad \rho>0,\quad \delta<0,\nonumber\\
&&WEC_{x}\quad\quad :\quad\quad \rho+p_{x}=0,\nonumber\\
&&WEC_{y}\quad\quad :\quad\quad \rho+p_{y}>0,\nonumber\\
&&WEC_{z}\quad\quad :\quad\quad \rho+p_{z}>0,\nonumber\\
&&SEC\quad\quad\quad:\quad\quad \rho+\sum_{i} p_{i}<0.
\label{m4}
\end{eqnarray}
While for $\delta>0$, we have
\begin{eqnarray}
&&WEC\quad\quad\quad:\quad\quad \rho<0,\nonumber\\
&&WEC_{x}\quad\quad :\quad\quad \rho+p_{x}=0,\nonumber\\
&&WEC_{y}\quad\quad :\quad\quad \rho+p_{y}<0,\nonumber\\
&&WEC_{z}\quad\quad :\quad\quad \rho+p_{z}<0,\nonumber\\
&&SEC\quad\quad\quad:\quad\quad \rho+\sum_{i} p_{i}>0.
\label{m5}
\end{eqnarray}

It is worth to mentioning that the conformally flat condition for static and cylindrically symmetric case with anisotropic fluids was obtained by Herrera {\it et al} \cite{Herr1}. They have shown there that any conformally flat and cylindrically symmetric static source cannot be matched through Darmois conditions to the Levi-Civita space-time, satisfying the regularity conditions. Furthermore, all static, cylindrical symmetry solutions (conformally flat or not) for anisotropic fluids have been found in \cite{Herr2}. Recently, two of the authors constructed a conformally flat cylindrical symmetry and static anisotropic solution of the Einstein's field equations with naked singularities \cite{Faiz6}. The study space-time (\ref{m1}) is a special case of these known solutions.

{ One may study whether the given space-time is geodesically complete or not. In general incompleteness of geodesic equation indicates the presence of a singularity, and if it is not clothed or covered by an event horizon, then the singularity is called naked singularity otherwise a black holes solution. For the metrics study in {\it Case 1} to {\it Case 4}, sub-case (ii) of {\it Case 7}, and the general metric in {\it Case 11}, the geodesic equations considering $y=const$ and $z=const$ are
\begin{eqnarray}
\ddot{t}&=&-\frac{f'(x)}{f(x)}\,\dot{x}\,\dot{t}\nonumber\\
\ddot{x}&=&\frac{1}{2}\,[\frac{f'(x)}{f(x)}\,\dot{x}^2-f'(x)\,f(x)\,\dot{t}^2],
\label{geodesics}
\end{eqnarray}
where dot stands derivative w. r. t. an affine parameter $s$ and prime w. r. t. $x$. Solving the above geodesic equations for $t$, its first derivative give
\begin{equation}
\dot{t}=\frac{c_1}{f(x)},
\label{geodesics2}
\end{equation}
where $c_1$ is a constant, $c_1\neq 0$. Substituting $\dot{t}$ in the above geodesic equations we get
\begin{equation}
\ddot{x}=\frac{1}{2}\,\frac{f'(x)}{f(x)}\,(\dot{x}^2-c^2_{1}).
\label{geodesics3}
\end{equation}
Its solution in first derivative gives
\begin{equation}
\dot{x} (s)=\sqrt{c^{2}_{1}+c_{2}\,f(x)},
\label{geodesics4}
\end{equation}
where $c_2$ is a constant, $c_2\neq 0$.

Below we consider {\it Case 4} to show incompleteness of the geodesics path. Using $f(x)=\delta\,x$ in Eq. (\ref{geodesics4}), we get
\begin{eqnarray}
&&\dot{x}(s)=\sqrt{c^{2}_{1}+c_{2}\,\delta\,x(s)}\nonumber\\\Rightarrow
&&\frac{2}{c_2\,\delta}\,\sqrt{c^{2}_{1}+c_{2}\,\delta\,x(s)}=(s+c_3)\nonumber\\ \Rightarrow
&&x(s)=\frac{1}{c_2\,\delta}\,[\frac{c^{2}_{2}\,\delta^2}{4}\,(s+c_3)^2-c^2_{1}].
\label{geodesics5}
\end{eqnarray}
If one choose the value of the affine parameter $s=0$, from (\ref{geodesics5}) we get
\begin{equation}
x(s=0)=\frac{1}{c_2\,\delta}\,[\frac{c^{2}_{2}\,c^{2}_{3}\,\delta^2}{4}-c^2_{1}]=0,
\label{geodesics6}
\end{equation}
where we set $c_3=\frac{2\,c_1}{c_2\,\delta}\neq 0$. Therefore from Eq. (\ref{geodesics2}) using $f(x)=\delta\,x$, we get
\begin{eqnarray}
\dot{t} (s)=\frac{c_1}{\delta\,x(s)}&=&\frac{c_1\,c_2}{[\frac{c^{2}_{2}\,\delta^2}{4}\,(s+c_3)^2-c^2_{1}]}\nonumber\\
&=&\frac{c_1\,c_2}{[\frac{c^{2}_{2}\,\delta^2}{4}\,(s+\frac{2\,c_1}{c_2\,\delta})^2-c^2_{1}]}
\label{geodesics7}
\end{eqnarray}
which is unbounded at $s=0$. Hence the geodesic path $\dot{t}$ is incomplete and the space-time is geodesically incomplete. Since the solution is neither a black hole model nor exist an event horizon. Therefore the solution possess a naked singularity.

Similarly, one can easily check using the function $f(x)$ given in {\it Case 1} to {\it Case 3} that the first derivative $\dot{t}$ is unbounded/infinite for finite values of the affine parameter including $s=0$, and so the space-time including the general metric in {\it Case 11} is geodesically incomplete}.

\vspace{0.1cm}
\begin{center}
{\bf Case 5} : $\beta\neq 0$, $\alpha=0$, $\Lambda\neq 0$, $\delta=0$, $\gamma\neq 0$.
\end{center}


The scalar curvature invariant is given by
\begin{eqnarray}
R_{\mu\nu\rho\sigma}\,R^{\mu\nu\rho\sigma}&=&\frac{48\,\beta^2}{x^6}-\frac{8\,\gamma^2}{x^2}+\frac{108\,x^2\,\gamma^4}{(-6\,\beta+\Lambda\,x^3)^2}\nonumber\\
&&+\frac{8}{3}[\Lambda^2-\frac{54\,\beta\,\gamma^2\,(3\,\beta+x^3\,\Lambda)}{(-6\,\beta\,x+x^4\,\Lambda)^2}].
\label{case5-scalar}
\end{eqnarray}
From above one can easily check that the curvature scalar diverge at $x=0$ and approaches $\frac{8\,\Lambda^2}{3}$ at $x\rightarrow \pm\,\infty$.

Now, we are going to discuss the region $x>0$ and $x<0$, respectively in this space-time.

In the region $x>0$, two possibilites arise:

\vspace{0.1cm}
\underline{\it Sub-case} (i): Considering $\beta\rightarrow -\beta$ and $\Lambda\rightarrow -\Lambda$.
\vspace{0.1cm}

The function $f(x)$ and the quantity $k(x)$ under this case
\begin{eqnarray}
f(x)&=&(\frac{\Lambda}{3}\,x^2-\frac{2\,\beta}{x})=\frac{2\,\beta}{x}\,(\frac{x^{3}}{x^{3}_{0}}-1),\nonumber\\
k(x)&=&(\frac{\Lambda}{3}\,x+\frac{\beta}{x^2})=\beta\,x\,(\frac{2}{x^{3}_{0}}+\frac{1}{x^3}).
\label{n1}
\end{eqnarray}
The function $f(x)$ is space-like in the regions $x > x_0$ and time-like in $0 < x < x_0$. The quantity $k(x)$ for $f(x_0)=0$ is given by $k(x_0)=\frac{\Lambda}{2}\,x_0$ which is space-like where, $x_0=(\frac{6\,\beta}{\Lambda})^{\frac{1}{3}}$. The metric under this case is asymptotically anti-de Sitter (AdS) spaces.

From the field equations using (\ref{perfect}) and (\ref{4}), we get
\begin{equation}
\kappa\,\rho=\kappa\,p=-\frac{\gamma^2}{\frac{2\,\beta}{x}\,(\frac{x^{3}}{x^{3}_{0}}-1)}.
\label{n2}
\end{equation}
The matter-energy content satisfy the different energy conditions in the regions $x < x_0$ and violate in $x>x_0$.

To show that the region $x \leq x_0$ represent a black hole region for the metric (\ref{1}) under this case with an event horizon exist at $x=x_0$, we apply the theorem (theorem 1.2.5) \cite{PT}. For that we do the transformation
\begin{equation}
dt\rightarrow dv-\frac{dx}{f(x)},
\label{n3}
\end{equation}
into the metric (\ref{1}), we get
\begin{eqnarray}
ds^2&=&-f(x)\,dv^2+2\,dv\,dx+x^2\,[H(v,x)\,dy^2+\frac{1}{H(v,x)}\,dz^2]\nonumber\\
&=&-g_{vv}\,dv^2+2\,dv\,dx+g_{yy}\,dy^2+g_{zz}\,dz^2,
\label{n4}
\end{eqnarray}
where
\begin{eqnarray}
H(v,x)&=&e^{-2\,\gamma\,v}\,e^{2\,\gamma\,\int\frac{dx}{f(x)}},\quad g_{vv}=f(x),\quad g_{yy}=x^2\,H(v,x),\nonumber\\
g_{vx}&=&1=g_{xv},\quad g_{zz}=x^2\,H^{-1}(v,x).
\label{n5}
\end{eqnarray}
The inverse metric tensor for the metric (\ref{n4}) can be express as
\begin{eqnarray}
g^{\mu\nu}\,\partial_{\mu}\,\partial_{\nu}&=&2\,\partial_{v}\,\partial_{x}+f(x)\,\partial_{x}^2+x^{-2}\,H^{-1}(v,x)\,\partial_{y}^2\nonumber\\
&+&x^{-2}\,H(v,x)\,\partial_{z}^2.
\label{n6}
\end{eqnarray}
As the metric component $g^{vv}$ from (\ref{n4}) is
\begin{equation}
g^{vv}=0\Rightarrow g(\nabla v,\nabla v)=0,
\label{n7}
\end{equation}
which implies that the integral curves of
\begin{equation}
\nabla v=g^{\mu\nu}\,\partial_{\mu}\,v\,\partial_{\nu}=\partial_{x},
\label{n8}
\end{equation}
are null, affinely parameterised geodesics, that means, the curves are the null geodesics (since the conditions $X^{\nu}\,\nabla_{\nu}\,X^{\mu}=0$ hold where, we have defined $X=\nabla\,v$).

From the metric (\ref{n4}), we have
\begin{equation}
\nabla x=g^{\mu\nu}\,\partial_{\mu}\,x\,\partial_{\nu}=\partial_{v}+f(x)\,\partial_{x}=\partial_{v}
\label{n9}
\end{equation}
provided $f(x)=0\Rightarrow x=x_0$. Since at $x=x_0$, the metric function $f(x)=0$ which implies $g^{xx}=0$, the null hypersurface condition. Therefore the curves $\gamma (\lambda)$ defined by $\{v=v(\lambda),\quad x=x_0,\quad y=y_0=const,\quad z=z_0=const\}$ are the null geodesic where, $\lambda$ is an affine parameter. They are, that is, the null geodesics are the generators of the event horizon. Thus the solution represent a black hole model of stiff fluid in the background of anti-de-Sitter (AdS) spaces.

Again let $\gamma(s)=(v(s), x(s), y(s), z(s))$ be a future directed time-like curve. From the metric (\ref{n4}) the condition $g(\dot{\gamma},\dot{\gamma})<0$ gives
\begin{equation}
-f(x)\,\dot{v}^2+2\,\dot{v}\,\dot{x}+x^2\,[H(v,x)\,\dot{y}^2+H^{-1}(v,x)\,\dot{z}^2]<0.
\label{n10}
\end{equation}
This implies that
\begin{equation}
\dot{v}\,(-f(x)\,\dot{v}+2\,\dot{x})<0,
\label{n11}
\end{equation}
where we have set $y=y_0$ and $z=z_0$ where, $y_0$, $z_0$ are constants. It follows that $\dot{v}$ does not change sign on future directed time-like curves. Since $\dot{v}>0$ in the region $x > x_0$ (from the standard choice of time orientation), which leads to
\begin{equation}
(-f(x)\,\dot{v}+2\,\dot{x})<0.
\label{n12}
\end{equation}
For $x \leq x_0$ region, the first term is non-negative which enforces $\dot{x}<0$ on all future directed time-like curves in that region. Thus, $x$ is a decreasing function along such curves, which implies that future directed time-like curves would cross the hypersurface $\{x=x_0=(\frac{6\,\beta}{\Lambda})^{\frac{1}{3}}\}$ if coming from the region $\{x > x_0\}$.

{ For light-like curves, from the metric (\ref{n4}) the null condition implies that ($y=const$, $z=const$)
\begin{equation}
-f(x)\,dv+2\,dx=0\Rightarrow 2\,\frac{dx}{dv}=f(x).
\label{n13}
\end{equation}
\quad Thus we see that
\begin{eqnarray}
\frac{dx}{dv}=\frac{1}{2}\,f(x)=\frac{\beta}{x}\,(\frac{x^3}{x^{3}_{0}}-1) &>& 0\quad \mbox{for}\quad x >x_0\nonumber\\
&<& 0\quad \mbox{for}\quad x <x_0
\label{n14}
\end{eqnarray}

Consider a congruence of null geodesics with tangent $k^{\mu}=\frac{dx^{\mu}}{d\lambda}$, where $\lambda$ is an affine parameter and satisfy the following condition
\begin{equation}
k^{\mu}\,k_{\mu}=0\quad,\quad k^{\nu}\,\nabla_{\nu}\,k^{\mu}=0.
\label{n15}
\end{equation}
Picking another null vector field $l^{\mu}$, such that $l^{\mu}\,l_{\mu}=0$ and $k^{\mu}\,l_{\mu}=-1$. The purely spatial metric $h_{\mu\nu}$, in the two-space orthogonal to $k^{\mu}$ is defined by
\begin{equation}
h_{\mu\nu}=g_{\mu\nu}+k_{\mu}\,l_{\nu}+l_{\mu}\,k_{\nu},
\label{n16}
\end{equation}
where $h_{\mu\nu}\,k^{\nu}=h_{\mu\nu}\,l^{\nu}=0$, $h^{\mu}_{\mu}=2$ and $h^{\mu}_{\nu}$ is the projection operator.

Let us consider a closed two-surface, $S$, of constant $v$ and $x$, from the metric (\ref{n4}) we have two null vector fields
\begin{eqnarray}
&&k^{\mu}=-\partial_{x}\quad \mbox{(the inner null normal)}\nonumber\\
&&l^{\mu}=\partial_{v}+\frac{1}{2}\,f(x)\,\partial_{x}\,\mbox{(the outer null normal)}.
\label{n17}
\end{eqnarray}
We compute the expansion scalar associated with ${\bf k, l}$ for two-surface \cite{Nielsen,Valerio}
\begin{eqnarray}
\Theta_{k}=h^{\mu\nu}\,\nabla_{\nu}\,k_{\mu}&=&-\frac{2}{x}\nonumber\\
\Theta_{l}=h^{\mu\nu}\,\nabla_{\nu}\,l_{\mu}&=&\frac{f(x)}{x}=\frac{2\,\beta}{x^2}\,(\frac{x^3}{x^{3}_{0}}-1).
\label{n18}
\end{eqnarray}
Once we have the expansions, following Penrose \cite{Roger-Penrose} we can define that the two-surface, $S$, of constant $v$ and $x$ is {\it trapped}, {\it marginally trapped}, or {\it untrapped}, according to whether $\Theta_{l}\,\Theta_{k}>0$, $\Theta_{l}\,\Theta_{k}=0$, or $\Theta_{l}\,\Theta_{k}<0$. An {\it apparent horizon}, or {\it trapping horizon} in Hayward's terminology \cite{Hay,Hay2} is defined as a hypersurface foliated by {\it marginally trapped} surfaces. In addition, in regions of mild curvature or a {\it normal} surface corresponds to $\Theta_{l}>0$ and $\Theta_{k}<0$. However, this need not be the case in regions of strong curvature.

Thus in our case
\begin{eqnarray}
\Theta_{l} &>&0\quad \mbox{for} \quad x > x_0\nonumber\\
&=&0\quad \mbox{for} \quad x = x_0\nonumber\\
&<&0\quad \mbox{for} \quad x < x_0
\label{n19}
\end{eqnarray}
\quad Thus, in the quotient manifold with metric $g=-f(x)\,dv^2+2\,dv\,dx$, we have
\begin{eqnarray}
\mbox{Regular region}, {\it R}&=&\{(v,x): x >x_0\},\nonumber\\
\mbox{Trapped region}, {\it T}&=&\{(v,x): x <x_0\},\nonumber\\
\mbox{Apparent region}, {\it A}&=&\{(v,x): x =x_0\}.
\label{n20}
\end{eqnarray}
The induced metric on {\it A} is
\begin{equation}
h=2\,dv\,dx=2\,(\frac{1}{2}\,f(x)\,dv)\,dv=f(x)\,dv^2=0,
\label{n21}
\end{equation}
a null hypersurface condition, where we have used Eq. (\ref{n14}). So {\it A} is a null hypersurface.

The 2-surface $S$ with $\Theta_{k}<0$ \cite{Nielsen,Valerio,Roger-Penrose,Hay2,Blau}
\begin{eqnarray}
S\quad \mbox{is called untrapped} \quad \mbox{if} \quad \Theta_{l}&>&0\nonumber\\
\quad\mbox{marginally trapped} \quad \mbox{if} \quad \Theta_{l}&=&0\nonumber\\
\quad\mbox{trapped} \quad \mbox{if} \quad \Theta_{l}&<&0
\label{n22}
\end{eqnarray}
and we can rephrase the above result as the statement
\begin{eqnarray}
S_{v,x}\quad \mbox{is untrapped} \quad \mbox{for} \quad x &>& x_0\nonumber\\
\quad\mbox{marginally trapped} \quad \mbox{for} \quad x &=& x_0\nonumber\\
\quad\mbox{trapped} \quad \mbox{for} \quad x &<& x_0
\label{n23}
\end{eqnarray}
A {\it future outer trapping horizon} (FOTH) is the closure of a surface foliated by {\it marginal surfaces}, such that $k^{\mu}\,\nabla_{\mu}\,\Theta_{l}<0$, $\Theta_{l}=0$ and $\Theta_{k}<0$. In our case, we have}
\begin{equation}
k^{\mu}\,\nabla_{\mu}\,\Theta_{l}=-2\,\beta\,(\frac{2}{x^3}+\frac{1}{x^{3}_{0}})<0,
\label{n24}
\end{equation}
{ since $\beta>0$.

Thus from above we have seen that the solution represent a black holes model in the backgrounds of anti-de Sitter (AdS) spaces with black holes region $x\leq x_0$}.

\vspace{0.1cm}
\underline{\it Sub-case} (ii): Considering $\beta>0$ and $\Lambda>0$.
\vspace{0.1cm}

The function $f(x)$ and the quantity $k(x)$ under this case
\begin{eqnarray}
f(x)&=&(\frac{2\,\beta}{x}-\frac{\Lambda}{3}\,x^2)=\frac{2\,\beta}{x}\,(1-\frac{x^{3}}{x^{3}_{0}})\nonumber\\
k(x)&=&-(\frac{\beta}{x^2}+\frac{\Lambda}{3}\,x)
\label{nn1}
\end{eqnarray}
From the field equations using (\ref{perfect}) and (\ref{4}), we get
\begin{equation}
\kappa\,\rho=\kappa\,p=-\frac{\gamma^2}{\frac{2\,\beta}{x}\,(1-\frac{x^{3}}{x^{3}_{0}})}.
\label{nn2}
\end{equation}
The space-time (\ref{1}) shows a true curvature singularity at $x=0$. The metric component $g^{xx}$ is given by
\begin{equation}
\quad g^{xx}=g(\nabla x,\nabla x)=f(x)=(\frac{2\,\beta}{x}-\frac{\Lambda}{3}\,x^2)=\frac{2\,\beta}{x}\,(1-\frac{x^3}{x^{3}_{0}}),
\label{nn3}
\end{equation}
which is time-like in the region $ x > x_0$ and space-like in $0 < x < x_0$. The matter-energy content stiff fluid satisfy the energy conditions in the region $ x> x_0$ and violate in $0 < x < x_0$.

{  Here for light-like curves, we have
\begin{eqnarray}
\frac{dx}{dv}=\frac{1}{2}\,f(x)=\frac{\beta}{x}\,(1-\frac{x^3}{x^{3}_{0}}) &>& 0\quad \mbox{for}\quad x <x_0\nonumber\\
&<& 0\quad \mbox{for}\quad x >x_0
\label{nn4}
\end{eqnarray}

We have obtained the expansions scalar as follows:
\begin{eqnarray}
\Theta_{k}=h^{\mu\nu}\,\nabla_{\nu}\,k_{\mu}&=&-\frac{2}{x},\nonumber\\
\Theta_{l}=h^{\mu\nu}\,\nabla_{\nu}\,l_{\mu}&=&\frac{f(x)}{x}=\frac{2\,\beta}{x^2}\,(1-\frac{x^3}{x^{3}_{0}}).
\label{nn5}
\end{eqnarray}
Thus in our case
\begin{eqnarray}
\Theta_{l} &>&0\quad \mbox{for} \quad x < x_0\nonumber\\
&=&0\quad \mbox{for} \quad x = x_0\nonumber\\
&<&0\quad \mbox{for} \quad x > x_0
\label{nn6}
\end{eqnarray}
For 2-surface $S$ with $\Theta_{k}<0$, the above result as the statement in the present case
\begin{eqnarray}
S_{v,x}\quad \mbox{is untrapped} \quad \mbox{for} \quad x &<& x_0\nonumber\\
\quad\mbox{marginally trapped} \quad \mbox{for} \quad x &=& x_0\nonumber\\
\quad\mbox{trapped} \quad \mbox{for} \quad x &>& x_0
\label{nn7}
\end{eqnarray}

Thus, in the quotient manifold with metric $g=-f(x)\,dv^2+2\,dv\,dx$, we have
\begin{eqnarray}
&&\mbox{Regular region}, {\it R}=\{(v,x): x <x_0\},\nonumber\\
&&\mbox{Trapped region}, {\it T}=\{(v,x): x >x_0\},\nonumber\\
&&\mbox{Apparent region}, {\it A}=\{(v,x): x =x_0\}\nonumber.
\end{eqnarray}
The induced metric on {\it A} is
\begin{equation}
h=2\,dv\,dx=2\,(\frac{1}{2}\,f(x)\,dv)\,dv=f(x)\,dv^2=0,
\label{nn8}
\end{equation}
a null hypersurface condition, where we have used Eq. (\ref{nn4}). So {\it A} is a null hypersurface.

From above analysis it is clear that the solution represent a black holes model in de-Sitter spaces background where the trapped region in $x > x_0$ and untrapped (regular) region in $x < x_0$}.

\vspace{0.1cm}
In the region $x<0$, two possibilites arise here.

\vspace{0.1cm}
\underline{\it Sub-case} (i): Considering $\Lambda>0$ and $\beta\rightarrow -\beta$
\vspace{0.1cm}

The function $f(x)$ and the quantity $k(x)$ under this case are
\begin{equation}
f(x)=(\frac{2\,\beta}{x}-\frac{\Lambda}{3}\,x^2),\quad k(x)=-(\frac{\beta}{x^2}+\frac{\Lambda}{3}\,x).
\label{nn3}
\end{equation}
Setting $f(x)=0$ implies $x=x_0=(\frac{6\,\beta}{\Lambda})^{\frac{1}{3}}>0$. Therefore the quantity $k(x)=-\frac{\Lambda}{2}\,x_0<0$ is time-like. The space-time possess a true curvature singularity at $x=0$ which is covered by an event horizon.

The physical parameters associated with the fluid
\begin{equation}
\kappa\,\rho=\kappa\,p=-\frac{\gamma^2}{(\frac{2\,\beta}{x}-\frac{\Lambda}{3}\,x^2)}.
\label{case-case}
\end{equation}
The energy-density of stress-energy tensor violate the energy conditions in the region $0 < x < x_0$, whereas obeying the energy conditions in the exterior region $x > x_0$. Thus the non-static solution represents a black holes model with stiff fluid as the sources and the solution is asymptotically de-Sitter spaces.

\vspace{0.1cm}
\underline{\it Sub-case} (ii): Considering $\Lambda\rightarrow -\Lambda$ and $\beta>0$
\vspace{0.1cm}

The function $f(x)$ and the quantity $k(x)$ under this case are
\begin{equation}
f(x)=(\frac{\Lambda}{3}\,x^2-\frac{2\,\beta}{x}),\quad k(x)=(\frac{\beta}{x^2}+\frac{\Lambda}{3}\,x).
\label{nn4}
\end{equation}
Setting $f(x)=0$ implies $x=x_0=(\frac{6\,\beta}{\Lambda})^{\frac{1}{3}}$. Therefore the quantity $k(x_0)=\frac{\Lambda}{2}\,x_0>0$ is space-like. The space-time possess a true curvature singularity at $x=0$ covered by an event hrizon.

The physical parameters associated with the fluid in the region $x<0$ are
\begin{equation}
\kappa\,\rho=\kappa\,p=-\frac{\gamma^2}{(\frac{\Lambda}{3}\,x^2-\frac{2\,\beta}{x})}.
\label{case-case2}
\end{equation}
The energy-density satisfy the energy conditions in the region $0< x < x_0$, whereas it violate in the region $x > x_0$. Thus the solution represent black holes model with stiff fluid as the stress-energy tensor and the solution is asymptotically anti-de-Sitter (AdS) spaces.

\vspace{0.1cm}
\begin{center}
{\bf Case 6} : $\beta=0$, $\alpha \neq 0$, $\Lambda=0$, $\gamma=0$, $\delta<0$.
\end{center}

The space-time under this case reduces to a static solution
\begin{equation}
ds^2=-f(x)\,dt^2+f^{-1}(x)\,dx^2+x^2\,(dy^2+dz^2),
\label{case6}
\end{equation}
where the function $f(x)$ and the quantity $k(x)$ are
\begin{equation}
f(x)=(\frac{\alpha^2}{x^2}+\delta\,x),\quad k(x)=(\frac{\delta}{2}-\frac{\alpha^2}{x^3}).
\label{nn5}
\end{equation}
Setting $f(x_0)=0$ which implies $x=x_0=(\frac{\alpha^2}{\delta_0})^{1/3}$, where $\delta=-\delta_0$ for $\delta_0>0$. Therefore the quantity $k(x_0)=-\frac{3}{2}\,\delta_0$ is time-like. Note that in the region $0 < x < x_0$, the function $f(x)>0$, therefore the coordinate $t$ is time-like and $x$ is space-like and vice-versa in the region $x > x_0$.

The scalar curvature invariant is given by
\begin{equation}
R^{\mu\nu\rho\sigma}\,R_{\mu\nu\rho\sigma}=\frac{8\,(7\,\alpha^4-x^3\,\alpha^2\,\delta+x^6\,\delta^2)}{x^8},
\label{nn6}
\end{equation}
From above it is clear that the scalar curvature diverge at $x=0$ clothed by an event horizon $x=x_0$. By analysing the same procedure as done earlier, one can easily show that the present solution represent a static charged black holes model.

Considering the stress-energy tensor the electromagnetic field coupled with anisotropic fluid (\ref{anisotropic}), from the field equations using (\ref{4}) we get
\begin{equation}
B=\frac{\alpha}{\sqrt{2}},\quad \kappa\,\rho=\frac{2\,\delta_0}{x}=-\kappa\,p_{x},\quad \kappa\,p_{y}=\kappa\,p_{z}=-\frac{\delta_0}{x},
\label{nn7}
\end{equation}
where the electromagnetic field tensor $F_{32}=-F_{23}=B^1=B$ such that the electromagnetic EMT is
\begin{equation}
-T^{0}_{0}=-T^{1}_{1}=T^{2}_{2}=T^{3}_{3}=\frac{2}{\kappa}\times\frac{B^2}{x^4}.
\label{nn8}
\end{equation}
The energy density of anisotropic fluid is positive in the region $x>0$, where the pressure components are negative. The matter-energy source satisfy the following energy conditions (\ref{energy-conditions}) in region $x>0$ only.
\begin{eqnarray}
&&WEC\quad\quad:\quad\quad \rho>0,\nonumber\\
&&WEC_{x}\quad\quad:\quad\quad \rho+p_{x}=0,\nonumber\\
&&WEC_{y}\quad\quad:\quad\quad \rho+p_{y}>0,\nonumber\\
&&WEC_{z}\quad\quad:\quad\quad \rho+p_{z}>0,\nonumber\\
&&SEC\quad\quad:\quad\quad \rho+\sum_{i} p_{i}<0.
\label{case6-energy}
\end{eqnarray}

\vspace{0.1cm}
\begin{center}
{\bf Case 7} : $\beta=0$, $\alpha \neq 0$, $\Lambda>0$, $\gamma\neq 0$, $\delta=0$.
\end{center}

The function $f(x)$ and the quantity $k(x)$ under this case are
\begin{equation}
f(x)=(\frac{\alpha^2}{x^2}-\frac{\Lambda}{3}\,x^2),\quad k(x)=-(\frac{\alpha^2}{x^3}+\frac{\Lambda}{3}\,x).
\label{nn9}
\end{equation}

The scalar curvature invariant is given by
\begin{eqnarray}
R^{\mu\nu\rho\sigma}\,R_{\mu\nu\rho\sigma}&=&8\,[\frac{7\,\alpha^4}{x^8}-\frac{\gamma^2}{x^2}+\frac{\Lambda^2}{3}]+\frac{108\,\gamma^4}{x^4\,(-\frac{3\,\alpha^2}{x^4}+\Lambda)^2}\nonumber\\
&&-\frac{96\,\alpha^2\,\gamma^2\,(\frac{3\,\alpha^2}{x^4}+\Lambda)}{x^6\,(-\frac{3\,\alpha^2}{x^4}+\Lambda)^2}.
\label{case7-scalar-curvature}
\end{eqnarray}
From above it is clear that there is a true curvature singularity at $x=0$ and a coordinate singularity at $x=x_0=(\frac{3\,\alpha^2}{\Lambda})^{\frac{1}{4}}$ which can be remove by the following transformations:
\begin{equation}
t\rightarrow v+\frac{x_{0}^{3}}{4\,\alpha^2}\,[2\,\mbox{tan}^{-1} (\frac{x}{x_0})+\mbox{ln}(\frac{x-x_0}{x+x_0})],
\label{nn10}
\end{equation}
we get
\begin{equation}
ds^2=-f(x)\,dv^2+2\,dv\,dx+x^2\,[A(v,x)\,dy^2+\frac{1}{A(v,x)}\,dz^2],
\label{nn11}
\end{equation}
where
\begin{eqnarray}
A(v,x)&=&e^{-2\,\gamma\,v} \nonumber\\
&&\times  e^{-\gamma\,\frac{x_{0}^{3}}{2\,\alpha^2}\,[2\,\mbox{tan}^{-1} (\frac{x}{x_0})+\mbox{ln}(\frac{x-x_0}{x+x_0})]}.
\label{nn12}
\end{eqnarray}

Choosing the electromagnetic field tensor $F_{32}=-F_{23}=B^1=B$ such that the electromagnetic EMT (\ref{em}) is
\begin{equation}
-T^{0}_{0}=-T^{1}_{1}=T^{2}_{2}=T^{3}_{3}=\frac{2}{\kappa}\times\frac{B^2}{x^4}.
\label{nn13}
\end{equation}

Considering the stress-energy tensor the electromagnetic field coupled with perfect fluid (\ref{perfect}), from the field equations using (\ref{4}) we get
\begin{equation}
B=\frac{\alpha}{\sqrt{2}},\quad \kappa\,\rho=\kappa\,p=\frac{\gamma^2}{(-\frac{\alpha^2}{x^2}+\frac{\Lambda}{3}\,x^2)},\quad \Lambda>0.
\label{nn14}
\end{equation}
We have seen that in the region $x < x_0$, $t$ is time-like and $x$ is a space-like coordinate. The energy-density as well as pressures in this region are negative violating the different energy conditions. On the other hand, in the region $x > x_0$, $t$ is space-like and $x$ is time-like coordinate. The energy density as well as the isotropic pressure are positive satisfying the different energy conditions. In addition, at spatial infinity along $x$-direction {\it i.e.,} $x\rightarrow \pm\,\infty$, the space-time is asymptotically de-Sitter space (since $R^{\mu\nu\rho\sigma}\,R_{\mu\nu\rho\sigma} \rightarrow \frac{8\,\Lambda^2}{3}$). Furthermore, the non-zero Weyl scalars are $\Psi_0\neq 0$, $\Psi_2\neq 0$ and $\Psi_4\neq 0$. By analysing the same procedure done earlier one can easily show that $x=x_0=(\frac{3\,\alpha^2}{\Lambda})^{\frac{1}{4}}$ represent a black holes event horizon. Therefore the studied metric is an algebraically general (Petrov type I) non-static charged black holes solution of the Einstein's field equations. Hence the solution with stress-energy tensor stiff fluid coupled with a non-null electromagnetic field (since $F^{\mu\nu}\,F_{\mu\nu}\neq 0$) in the background of de-Sitter spaces (dS) represent a non-static charged black holes model.

Now we will discuss two sub-cases as below:

\underline{\it Sub-case} (i): If one takes $\gamma=0$, then the non-static solution reduces to static one given by
\begin{equation}
ds^2=-(\frac{\alpha^2}{x^2}-\frac{\Lambda}{3}\,x^2)\,dt^2+f^{-1}(x)\,dx^2+x^2\,(dy^2+dz^2).
\label{nn15}
\end{equation}
The matter-energy source corresponds to non-null electromagnetic field in the background of de-Sitter Universe. The surface gravity $k(x)$ for $f(x_0)=0$ is $k(x_0)=-\frac{2\,\Lambda}{3}\,x_0$ which is time-like where, $x_0=(\frac{3\,\alpha^2}{\Lambda})^{\frac{1}{4}}$ (since $\Lambda>0$). To overcome the coordinate singularity at $x=x_0=(\frac{3\,\alpha^2}{\Lambda})^{\frac{1}{4}}$, we do the following transformation
\begin{equation}
t\rightarrow v+\frac{x_{0}^{3}}{4\,\alpha^2}\,[2\,\mbox{tan}^{-1} (\frac{x}{x_0})+\mbox{ln} (\frac{x-x_0}{x+x_0})],
\label{nn16}
\end{equation}
into the metric (\ref{1}), we get a static charged balck holes given by
\begin{equation}
ds^2=-(\frac{\alpha^2}{x^2}-\frac{\Lambda}{3}\,x^2)\,dv^2+2\,dv\,dx+x^2\,(dy^2+dz^2),
\label{nn17}
\end{equation}
which can also be obtain directly from (\ref{nn11}) and (\ref{nn12}) by substituting $\gamma=0$.

The scalar curvature invariant (\ref{case7-scalar-curvature}) given by
\begin{equation}
R^{\mu\nu\rho\sigma}\,R_{\mu\nu\rho\sigma}=\frac{8\,\Lambda^2}{3}+\frac{56\,\alpha^4}{x^8},
\label{nn18}
\end{equation}
diverge at $x=0$. The true curvature singularity of the solution occurs at $x=0$ covered by an event horizon.

Therefore the type D solution with non-null electromagnetic field as the stress-energy tensor in the background of de-Sitter Universe represent a static charged black holes model.

\underline{\it Sub-case} (ii): If one takes $\Lambda=0$ and $\gamma=0$, then the static solution is given by
\begin{equation}
ds^2=-\frac{\alpha^2}{x^2}\,dt^2+\frac{x^2}{\alpha^2}\,dx^2+x^2\,(dy^2+dz^2).
\label{nn19}
\end{equation}
The above static solution corresponds to a non-null electromagnetic field with zero cosmological constant and the metric is of Petrov type D. By doing a transformation $t\rightarrow v-\frac{x^3}{3\,\alpha^2}$ into the above metric, one will get
\begin{equation}
ds^2=-\frac{\alpha^2}{x^2}\,dv^2+2\,dv\,dx+x^2\,(dy^2+dz^2).
\label{nn20}
\end{equation}

The scalar curvature invariant (\ref{case7-scalar-curvature}) given by
\begin{equation}
R^{\mu\nu\rho\sigma}\,R_{\mu\nu\rho\sigma}=\frac{56\,\alpha^4}{x^8}.
\label{nn21}
\end{equation}
diverge at $x=0$. Thus the solution possess a true curvature singularity at $x=0$ which is not covered by an event horizon and therefore a naked singularity is formed in that sub-case. Thus the type D non-null electromagnetic field charged solution possess a naked singularity.

\vspace{0.1cm}
\begin{center}
{\bf Case 8} : $\beta\rightarrow-\beta$, $\alpha \neq 0$, $\Lambda=0$, $\gamma\neq 0$, $\delta=0$.
\end{center}

The function $f(x)$ and the quantity $k(x)$ under this case are
\begin{equation}
f(x)=(\frac{\alpha^2}{x^2}-\frac{2\,\beta}{x}),\quad k(x)=(-\frac{\alpha^2}{x^3}+\frac{\beta}{x^2}).
\label{nn22}
\end{equation}
For $f(x)=0$ which implies $x=x_0=\frac{\alpha^2}{2\,\beta}$, the quantity $k(x_0)=-\frac{4\,\beta^3}{\alpha^4}$, is time-like.

The scalar curvature invariant is given by
\begin{eqnarray}
R^{\mu\nu\rho\sigma}\,R_{\mu\nu\rho\sigma}&=&\frac{8}{x^8}\,(7\,\alpha^4-12\,x\,\alpha^2\,\beta+6\,x^2\,\beta^2)\nonumber\\
&&+\frac{12\,\gamma^4\,x^4}{(\alpha^2-2\,\beta\,x)^2}-\frac{8\,\gamma^2}{x^2\,(\alpha^2-2\,\beta\,x)^2}\nonumber\\
&&\times (5\,\alpha^4-14\,x\,\alpha^2\,\beta+10\,x^2\,\beta^2).
\label{nn23}
\end{eqnarray}
From above it is clear that there is a true curvature singularity at $x=0$ and a coordinate singularity at $x=x_0=\frac{\alpha^2}{2\,\beta}$ which can removable by suitable transformations.

Choosing the electromagnetic field tensor $F_{32}=-F_{23}=B^1=B$ such that the electromagnetic EMT (\ref{em}) is
\begin{equation}
-T^{0}_{0}=-T^{1}_{1}=T^{2}_{2}=T^{3}_{3}=\frac{2}{\kappa}\times\frac{B^2}{x^4}.
\label{nn24}
\end{equation}

Considering the stress-energy tensor electromagnetic field coupled with perfect fluid (\ref{perfect}), from the field equations using (\ref{4}), we get
\begin{equation}
B=\frac{\alpha}{\sqrt{2}}>0,\quad \kappa\,\rho=\kappa\,p=\frac{\gamma^2}{(\frac{2\,\beta}{x}-\frac{\alpha^2}{x^2})}.
\label{nn25}
\end{equation}
The perfect fluid which here is a stiff fluid satisfy the different energy conditions (\ref{energy-conditions}) in the region $x > x_0$, where $t$ is space-like and $x$ is time-like coordinate. On the other hand, the fluid violates the different energy conditions in the exterior region $0 < x < x_0$, where $t$ is time-like and $x$ is space-like coordinate.

If one choose $\gamma=0$, then the solution represents a static charged black hole space-time with non-null electromagnetic field as the source. The metric under this condition is given by
\begin{equation}
ds^2=-f(x)\,dt^2+f^{-1}(x)\,dx^2+x^2\,(dy^2+dz^2).
\label{case88}
\end{equation}
where $f(x)$ is given in (\ref{nn22}). By doing the following transformation
\begin{equation}
dt\rightarrow dv-\frac{dx}{f(x)}
\label{gg-ss}
\end{equation}
into the metric (\ref{case88}), we get
\begin{equation}
ds^2=-(\frac{\alpha^2}{x^2}-\frac{2\,\beta}{x})\,dv^2+2\,dv\,dx+x^2\,(dy^2+dz^2).
\label{case888}
\end{equation}
By analysing the same procedure done earlier one can easily show that $x=x_0=\frac{\alpha^2}{2\,\beta}$ represent a black holes event horizon. Thus the solution represent a static charged black holes model with non-null electromagnetic field.

\vspace{0.1cm}
\begin{center}
{\bf Case 9} : $\beta\neq 0$, $\alpha=0$, $\Lambda=0$, $\gamma\neq 0$, $\delta\neq 0$.
\end{center}
\vspace{0.1cm}

The function $f(x)$ and the quantity $k(x)$ under this case are
\begin{equation}
f(x)=(\frac{2\,\beta}{x}+\delta\,x),\quad k(x)=\frac{1}{2}\,(-\frac{2\,\beta}{x^2}+\delta).
\label{nn26}
\end{equation}

The scalar curvature invariant is given by
\begin{eqnarray}
R^{\mu\nu\rho\sigma}\,R_{\mu\nu\rho\sigma}&=&\frac{4}{x^6}\,[12\,\beta^2-x^4\,(\gamma^2-2\,\delta^2)\nonumber\\
&&+\frac{\gamma^2\,x^4\,(-16\,\beta^2+3\,x^4\,\gamma^2)}{(2\,\beta+\delta\,x^2)^2}].
\label{case9-scalar}
\end{eqnarray}

Considering the stress-energy tensor anisotropic fluid (\ref{anisotropic}), from the field equations using (\ref{4}) we get
\begin{eqnarray}
-\kappa\,\rho&=&\frac{2\,\delta}{x}+\frac{\gamma^2}{\frac{2\,\beta}{x}+\delta\,x},\quad \kappa\,p_{x}=\frac{2\,\delta}{x}-\frac{\gamma^2}{\frac{2\,\beta}{x}+\delta\,x},\nonumber\\
\kappa\,p_{y}&=&\kappa\,p_{z}=\frac{\delta}{x}-\frac{\gamma^2}{\frac{2\,\beta}{x}+\delta\,x}.
\label{nn27}
\end{eqnarray}
The matter-energy content violate the weak energy condition in the region $x>0$.

Transforming $t\rightarrow v-\int \frac{dx}{f(x)}$ into the metric, we get
\begin{equation}
ds^2=-f(x)\,dv^2+2\,dv\,dx+x^2\,[H(v,x)\,dy^2+\frac{1}{H(v,x)}\,dz^2],
\label{case9-transfor}
\end{equation}
where
\begin{equation}
H(v,x)=e^{-2\,\gamma\,v}\,e^{2\,\gamma\,\int \frac{dx}{f(x)}}.
\label{mmnnmm}
\end{equation}

We will discuss the following sub-case:

\vspace{0.1cm}
\underline{\it Sub-case} (i): $\beta\rightarrow -\beta$,\quad $\delta>0$
\vspace{0.1cm}

The quantity $k(x)$ for $f(x_0)=0$ is $k(x_0)=\delta>0$ which is space-like, where $x_0=\sqrt{\frac{2\,\beta}{\delta}}$. There is a true curvature singularity at $x=0$, and a coordinate singularity at $x=\sqrt{\frac{2\,\beta}{\delta}}$.

The Physical parameters associated with anisotropic fluid are
\begin{eqnarray}
-\kappa\,\rho&=&\frac{2\,\delta}{x}+\frac{\gamma^2}{(\delta\,x-\frac{2\,\beta}{x})},\quad \kappa\,p_{x}=\frac{2\,\delta}{x}-\frac{\gamma^2}{(\delta\,x-\frac{2\,\beta}{x})},\nonumber\\
\kappa\,p_{y}&=&\kappa\,p_{z}=\frac{\delta}{x}-\frac{\gamma^2}{(\delta\,x-\frac{2\,\beta}{x})}.
\label{nn28}
\end{eqnarray}
The scalar curvature invariant from (\ref{case9-scalar}) diverge at $x=0$, a true curvature singularity and at $x=x_0$, a coordinate singularity which we can removable. From the physical parameters (\ref{nn28}), it is clear that in the region $x>0$, the matter-energy content anisotropic fluid violate the different energy conditions in the exterior region $x > x_0$. On the other hand, the matter-energy content satisfy the energy conditions in the region $0 < x < x_0$, that is, for $f(x) < 0$ provided the condition $\frac{\gamma^2}{(\delta\,x-\frac{2\,\beta}{x})}>\frac{2\,\delta}{x}$ hold.

By analysing the same procedure done earlier one can easily show that the surface $x=x_0=\sqrt{\frac{2\,\beta}{\delta}}$ represent a black holes event horizon. Thus the non-static solution with stress-energy tensor the anisotropic fluid represents an uncharged black holes model.

For $\gamma=0$, the solution (\ref{1}) under this case represents a static uncharged black holes model with zero cosmological constant given by
\begin{equation}
ds^2=-(\delta\,x-\frac{2\,\beta}{x})\,dv^2+2\,dv\,dx+x^2\,(dy^2+dz^2),
\label{nn29}
\end{equation}
with the physical parameters
\begin{equation}
-\kappa\,\rho=\frac{2\,\delta}{x},\quad \kappa\,p_{x}=\frac{2\,\delta}{x},\quad \kappa\,p_{y}=\kappa\,p_{z}=\frac{\delta}{x}.
\label{nn30}
\end{equation}
From above it is thus clear that the anisotropic fluid violate the weak energy conditions (WEC) in the region $x>0$ and satisfy in the region $x<0$. Also the solution is asymptotically flat and provide an example of a static uncharged black holes model.

\vspace{0.1cm}
\underline{\it Sub-case} (ii): $\beta>0$,\quad $\delta\rightarrow -\delta$
\vspace{0.1cm}

The quantity $k(x)$ for $f(x_0)=0$ is $k(x_0)=-\delta<0$ which is time-like where, $x=x_0=\sqrt{\frac{2\,\beta}{\delta}}$. There is a true curvature singularity at $x=0$, and a coordinate singularity occurs at $x=\sqrt{\frac{2\,\beta}{\delta}}$.

The physical parameters associated with anisotropic fluid are
\begin{eqnarray}
\kappa\,\rho&=&\frac{2\,\delta}{x}-\frac{\gamma^2}{\frac{2\,\beta}{x}-\delta\,x},\quad \kappa\,p_{x}=-\frac{2\,\delta}{x}-\frac{\gamma^2}{\frac{2\,\beta}{x}-\delta\,x},\nonumber\\
\kappa\,p_{y}&=&\kappa\,p_{z}=-\frac{\delta}{x}-\frac{\gamma^2}{\frac{2\,\beta}{x}-\delta\,x}.
\label{nn31}
\end{eqnarray}
The energy-density satisfy the weak energy-conditions in the region $x>0$ provided $x < x_0$.

By analysing the same procedure done earlier one can easily show that $x=x_0=\sqrt{\frac{2\,\beta}{\delta}}$ represent a black holes event horizon. Thus the non-static solution with stress-energy tensor anisotropic fluid represents an uncharged black holes model.

For $\gamma=0$, from the metric (\ref{1}) we get a static black hole space-time and the solution is asymptotically flat. The static black hole space-time with an event horizon $x=x_0=\sqrt{\frac{2\,\beta}{\delta}}$ is given by
\begin{equation}
ds^2=-(\frac{2\,\beta}{x}-\delta\,x)\,dv^2+2\,dv\,\,dx+x^2\,(dy^2+dz^2),
\label{nn32}
\end{equation}
with the physical parameters
\begin{equation}
\kappa\,\rho=\frac{2\,\delta}{x},\quad \kappa\,p_{x}=-\frac{2\,\delta}{x},\quad \kappa\,p_{y}=\kappa\,p_{z}=-\frac{\delta}{x}.
\label{nn33}
\end{equation}
The physical parameters satisfy the following conditions in the region $x>0$ as:
\begin{eqnarray}
&&WEC\quad\quad\quad : \quad\quad\quad \rho>0,\nonumber\\
&&WEC_{x}\quad\quad\quad :\quad\quad\quad \rho+p_{x}=0,\nonumber\\
&&WEC_{y}\quad\quad\quad :\quad\quad\quad \rho+p_{y}>0,\nonumber\\
&&WEC_{z}\quad\quad\quad :\quad\quad\quad \rho+p_{z}>0,\nonumber\\
&&SEC\quad\quad\quad\quad:\quad\quad\quad \rho+\sum_{i} p_{i}<0.
\label{case9-energy}
\end{eqnarray}
And in the region $x<0$, we have
\begin{eqnarray}
&&WEC\quad\quad\quad : \quad\quad\quad \rho<0,\nonumber\\
&&WEC_{x}\quad\quad\quad :\quad\quad\quad \rho+p_{x}=0,\nonumber\\
&&WEC_{y}\quad\quad\quad :\quad\quad\quad \rho+p_{y}<0,\nonumber\\
&&WEC_{z}\quad\quad\quad :\quad\quad\quad \rho+p_{z}<0,\nonumber\\
&&SEC\quad\quad\quad\quad:\quad\quad\quad \rho+\sum_{i} p_{i}<0.
\label{case9-energy2}
\end{eqnarray}

\vspace{0.1cm}
\begin{center}
{\bf Case 10} : $\beta=0$, $\alpha=0$, $\Lambda\neq 0$, $\gamma\neq 0$, $\delta\neq 0$.
\end{center}

The function $f(x)$ and the quantity $k(x)$ under this case are
\begin{equation}
f(x)=(\delta\,x-\frac{\Lambda}{3}\,x^2),\quad k(x)=\frac{1}{2}\,(\delta-\frac{2\,\Lambda}{3}\,x).
\label{nn34}
\end{equation}
One can easily show that there is a true curvature singularity at $x=0$ covered by an event horizon.

The scalar curvature invariant is given by
\begin{eqnarray}
R^{\mu\nu\rho\sigma}\,R_{\mu\nu\rho\sigma}&=&\frac{4}{x^2}\,[\frac{27\,\gamma^2}{(\Lambda\,x-3\,\delta)^2}
+\frac{2}{3}\,(3\,\delta^2-3\,\delta\,x\,\Lambda+\Lambda^2\,x^2)\nonumber\\
&-&\frac{\gamma^2}{(\Lambda\,x-3\,\delta)^2}\,(9\,\delta^2-6\,x\,\Lambda\,\delta+2\,x^2\,\Lambda^2)].
\label{case10-scalar}
\end{eqnarray}
From above we have seen that there is a true curvature singularity at $x=0$ covered by an event horizon. Note that there is coordinate singularity at $x=x_0=\frac{3\,\delta}{\Lambda}$. By analysing the same procedure done earlier one can easily show that the surface $x=x_0=\frac{3\,\delta}{\Lambda}$ represents the event horizon of an uncharged black holes solution.

Considering the stress-energy tensor anisotropic fluid (\ref{anisotropic}), from the field equations using (\ref{4}) we get
\begin{eqnarray}
&&\Lambda<0\quad \mbox{or}\quad >0,\nonumber\\
-\kappa\,\rho&=&\frac{2\,\delta}{x}+\frac{\gamma^2}{(\delta\,x-\frac{\Lambda}{3}\,x^2)}\nonumber\\
\kappa\,p_{x}&=&\frac{2\,\delta}{x}-\frac{\gamma^2}{(\delta\,x-\frac{\Lambda}{3}\,x^2)}\nonumber\\
\kappa\,p_{y}&=&\kappa\,p_{z}=\frac{\delta}{x}-\frac{\gamma^2}{\delta\,x-\frac{\Lambda}{3}\,x^2}.
\label{nn35}
\end{eqnarray}

There are two possibilites arise as below:

\vspace{0.1cm}
(i) $\delta>0$,\quad $\Lambda>0$
\vspace{0.1cm}

The quantity $k(x)$ for $f(x_0)=0$ is $k(x_0)=-\frac{\delta}{2}<0$ which is time-like where, $x_0=(\frac{3\,\delta}{\Lambda})$. One can easily by suitable transformation show that the coordinate singularity occurs at $x=(\frac{3\,\delta}{\Lambda})$ can easily be remove.

Transforming $t\rightarrow v+\frac{1}{\delta}\,\mbox{ln} (\frac{x}{\Lambda\,x-3\,\delta})$ into the metric, we get
\begin{eqnarray}
ds^2&=&-(\delta\,x-\frac{\Lambda}{3}\,x^2)\,dv^2+2\,dv\,dx\nonumber\\
     &&+x^2\,[H(v,x)\,dy^2+H(v,x)^{-1}\,dz^2],
\label{nn36}
\end{eqnarray}
where $H(v,x)=e^{-2\,\gamma\,v}\,(\frac{x}{\Lambda\,x-3\,\delta})^{-\frac{2\,\gamma}{\delta}}$.

For $\gamma=0$, the solution (\ref{nn36}) given by
\begin{equation}
ds^2=-(\delta\,x-\frac{\Lambda}{3}\,x^2)\,dv^2+2\,dv\,dx+x^2\,(dy^2+dz^2),
\label{nn37}
\end{equation}
represents a static uncharged black hole space-time with positive cosmological constant ($\Lambda>0$). The stress-energy tensor anisotropic fluid (\ref{nn35}) satisfy the following energy-conditions in the region $x<0$
\begin{eqnarray}
&&WEC\quad\quad\quad : \quad\quad\quad \rho>0,\nonumber\\
&&WEC_{x}\quad\quad\quad :\quad\quad\quad \rho+p_{x}=0,\nonumber\\
&&WEC_{y}\quad\quad\quad :\quad\quad\quad \rho+p_{y}>0,\nonumber\\
&&WEC_{z}\quad\quad\quad :\quad\quad\quad \rho+p_{z}>0,\nonumber\\
&&SEC\quad\quad\quad\quad:\quad\quad\quad \rho+\sum_{i} p_{i}>0.
\label{nn38}
\end{eqnarray}
And in the region $x>0$, we have
\begin{eqnarray}
&&WEC\quad\quad\quad : \quad\quad\quad \rho<0,\nonumber\\
&&WEC_{x}\quad\quad\quad :\quad\quad\quad \rho+p_{x}=0,\nonumber\\
&&WEC_{y}\quad\quad\quad :\quad\quad\quad \rho+p_{y}<0,\nonumber\\
&&WEC_{z}\quad\quad\quad :\quad\quad\quad \rho+p_{z}<0,\nonumber\\
&&SEC\quad\quad\quad\quad:\quad\quad\quad \rho+\sum_{i} p_{i}<0.
\label{nn39}
\end{eqnarray}
Therefore the solution represent a static uncharged black hole models in the background of de-Sitter (dS) spaces.

\vspace{0.1cm}
(ii) $\delta\rightarrow -\delta$,\quad $\Lambda\rightarrow -\Lambda$
\vspace{0.1cm}

The quantity $k(x)$ for $f(x_0)=0$ is $k(x_0)=\frac{\delta}{2}>0$ which is space-like where, $x_0=(\frac{3\,\delta}{\Lambda})$.  One can easily remove the coordinate singularity $x=(\frac{3\,\delta}{\Lambda})$ by suitable transformation.

Transforming $t\rightarrow v+\frac{1}{\delta}\,\mbox{ln} (\frac{x}{\Lambda\,x-3\,\delta})$ into the metric, one will get a non-static uncharged black holes solution given by
\begin{eqnarray}
ds^2&=&-(\frac{\Lambda}{3}\,x^2-\delta\,x)\,dv^2+2\,dv\,dx\nonumber\\
&&+x^2\,[M(v,x)\,dy^2+M(v,x)^{-1}\,dz^2],
\label{nn40}
\end{eqnarray}
where
\begin{equation}
M(v,x)=e^{-2\,\gamma\,v}\,(\frac{x}{\Lambda\,x-3\,\delta})^{-\frac{2\,\gamma}{\delta}}.
\label{nn41}
\end{equation}

The stress-energy tensor anisotropic fluid is given by
\begin{eqnarray}
-\kappa\,\rho&=&-\frac{2\,\delta}{x}+\frac{\gamma^2}{(\frac{\Lambda}{3}\,x^2-\delta\,x)}\nonumber\\
\kappa\,p_{x}&=&-\frac{2\,\delta}{x}-\frac{\gamma^2}{(\frac{\Lambda}{3}\,x^2-\delta\,x)}\nonumber\\
\kappa\,p_{y}&=&\kappa\,p_{z}=-\frac{\delta}{x}-\frac{\gamma^2}{(\frac{\Lambda}{3}\,x^2\,x^2-\delta\,x)}.
\label{nn42}
\end{eqnarray}

For $\gamma=0$, the solution (\ref{nn40}) reduces to
\begin{equation}
ds^2=-(\frac{\Lambda}{3}\,x^2-\delta\,x)\,dv^2+2\,dv\,dx+x^2\,(dy^2+dz^2),
\label{nn43}
\end{equation}
a static uncharged black hole space-time with negative cosmological constant. The stress-energy tensor anisotropic fluid satisfy the following energy-conditions in the region $x>0$
\begin{eqnarray}
&&WEC\quad\quad\quad : \quad\quad\quad \rho>0,\nonumber\\
&&WEC_{x}\quad\quad\quad :\quad\quad\quad \rho+p_{x}=0,\nonumber\\
&&WEC_{y}\quad\quad\quad :\quad\quad\quad \rho+p_{y}>0,\nonumber\\
&&WEC_{z}\quad\quad\quad :\quad\quad\quad \rho+p_{z}>0,\nonumber\\
&&SEC\quad\quad\quad\quad:\quad\quad\quad \rho+\sum_{i} p_{i}<0.
\label{nn44}
\end{eqnarray}
And in the region $x<0$, we have
\begin{eqnarray}
&&WEC\quad\quad\quad : \quad\quad\quad \rho<0,\nonumber\\
&&WEC_{x}\quad\quad\quad :\quad\quad\quad \rho+p_{x}=0,\nonumber\\
&&WEC_{y}\quad\quad\quad :\quad\quad\quad \rho+p_{y}<0,\nonumber\\
&&WEC_{z}\quad\quad\quad :\quad\quad\quad \rho+p_{z}<0,\nonumber\\
&&SEC\quad\quad\quad\quad:\quad\quad\quad \rho+\sum_{i} p_{i}>0.
\label{nn45}
\end{eqnarray}
The solution in that case represent a static black hole model in the background of anti-de Sitter (AdS) spaces.

\vspace{0.1cm}
\begin{center}
{\bf Case 11} : $\beta \neq 0$, $\alpha \neq 0$, $\Lambda \neq 0$, $\gamma\neq 0$, $\delta  \neq 0$.
\end{center}

In this case we take all the parameters to be non-vanishing. We discuss the naked singularities graphically, {\it i. e.}, one can see that no horizon exists for some specific values of the parameters. Figure 1 indicates that naked singularities exist for positive values of $\beta$ with both de-Sitter and anti de-Sitter spacetimes. Figure 2 indicates that naked singularities exist for negative values of $\beta$ with both de-Sitter and anti de-Sitter spacetimes. Figure 3 indicates that naked singulairites exist for anti-de Sitter spaces with both positive and negative values of $\beta$.

Choosing the electromagnetic field tensor $F_{32}=-F_{23}=B^1=B$ such that the electromagnetic EMT (\ref{em}) is
\begin{equation}
-T^{0}_{0}=-T^{1}_{1}=T^{2}_{2}=T^{3}_{3}=\frac{2}{\kappa}\times\frac{B^2}{x^4}.
\label{nn46}
\end{equation}

Choosing the stress-energy tensor non-null electromagnetic field coupled with anisotropic fluid, the physical parameters are
\begin{eqnarray}
B&=&\frac{\alpha}{\sqrt{2}},\quad \Lambda < 0\quad \mbox{or}\quad >0,\nonumber\\
\kappa\,\rho&=&\frac{3\,\gamma^2\,x^2}{-3\,(\alpha^2+2\,\beta\,x+\delta\,x^3)+x^4\,\Lambda}-\frac{2\,\delta}{x},\nonumber\\
\kappa\,p_{x}&=&\frac{3\,\gamma^2\,x^2}{-3\,(\alpha^2+2\,\beta\,x+\delta\,x^3)+x^4\,\Lambda}+\frac{2\,\delta}{x},\nonumber\\
\kappa\,p_{y}=\kappa\,p_{z}&=&\frac{3\,\gamma^2\,x^2}{-3\,(\alpha^2+2\,\beta\,x+\delta\,x^3)+x^4\,\Lambda}+\frac{\delta}{x}.
\label{final-case}
\end{eqnarray}
The ansiotropy difference along $x$-, $y$- and $z$-axis, respectively are
\begin{equation}
\kappa\,(\rho-p_{x})=-\frac{4\,\delta}{x},\quad \kappa\,(\rho-p_{y})=\kappa\,(\rho-p_{z})=-\frac{3\,\delta}{x}.
\label{final-case2}
\end{equation}

\begin{figure*}[tbp]
\centering
\includegraphics[width=6.5cm]{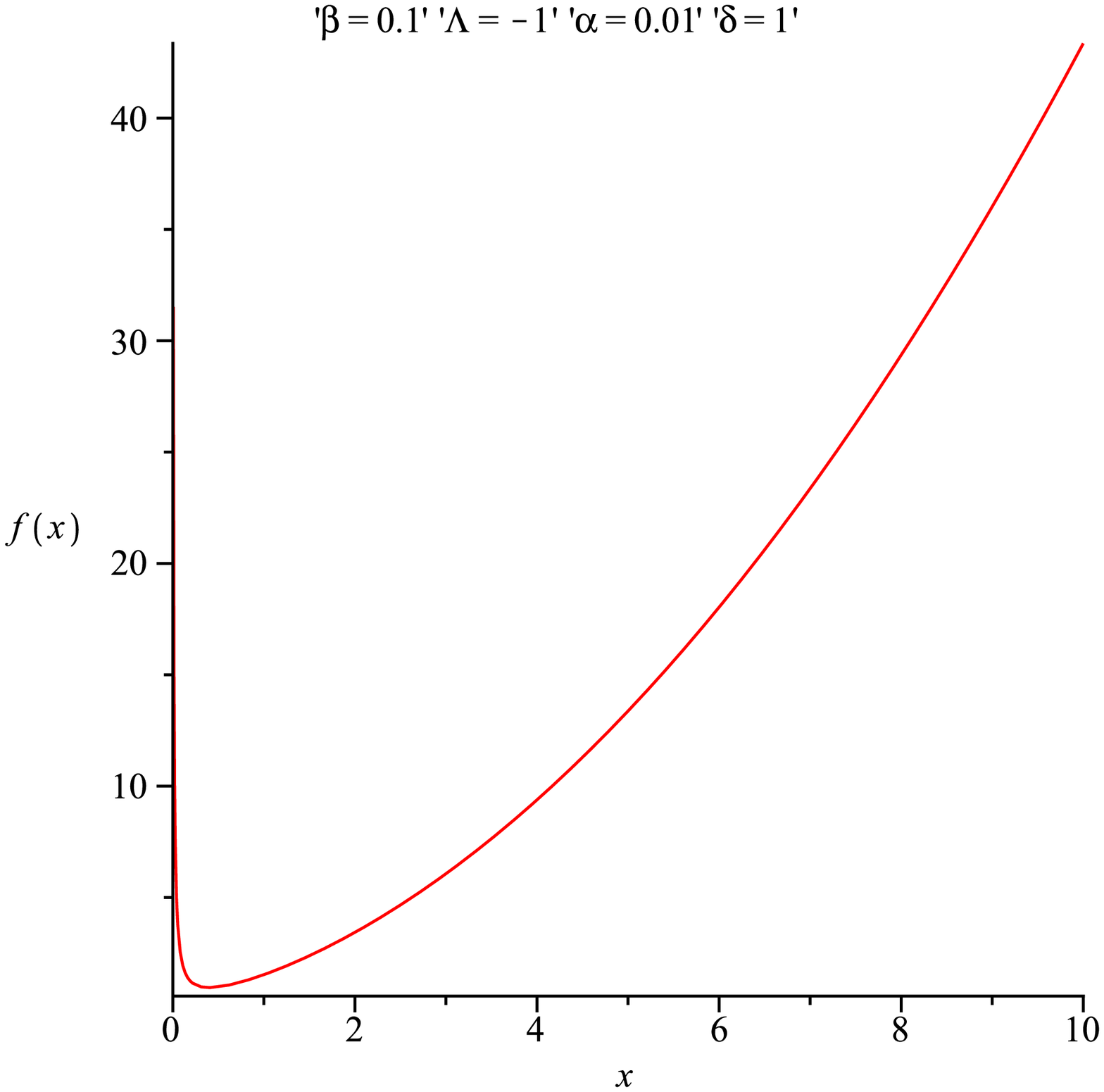}
\includegraphics[width=6.5cm]{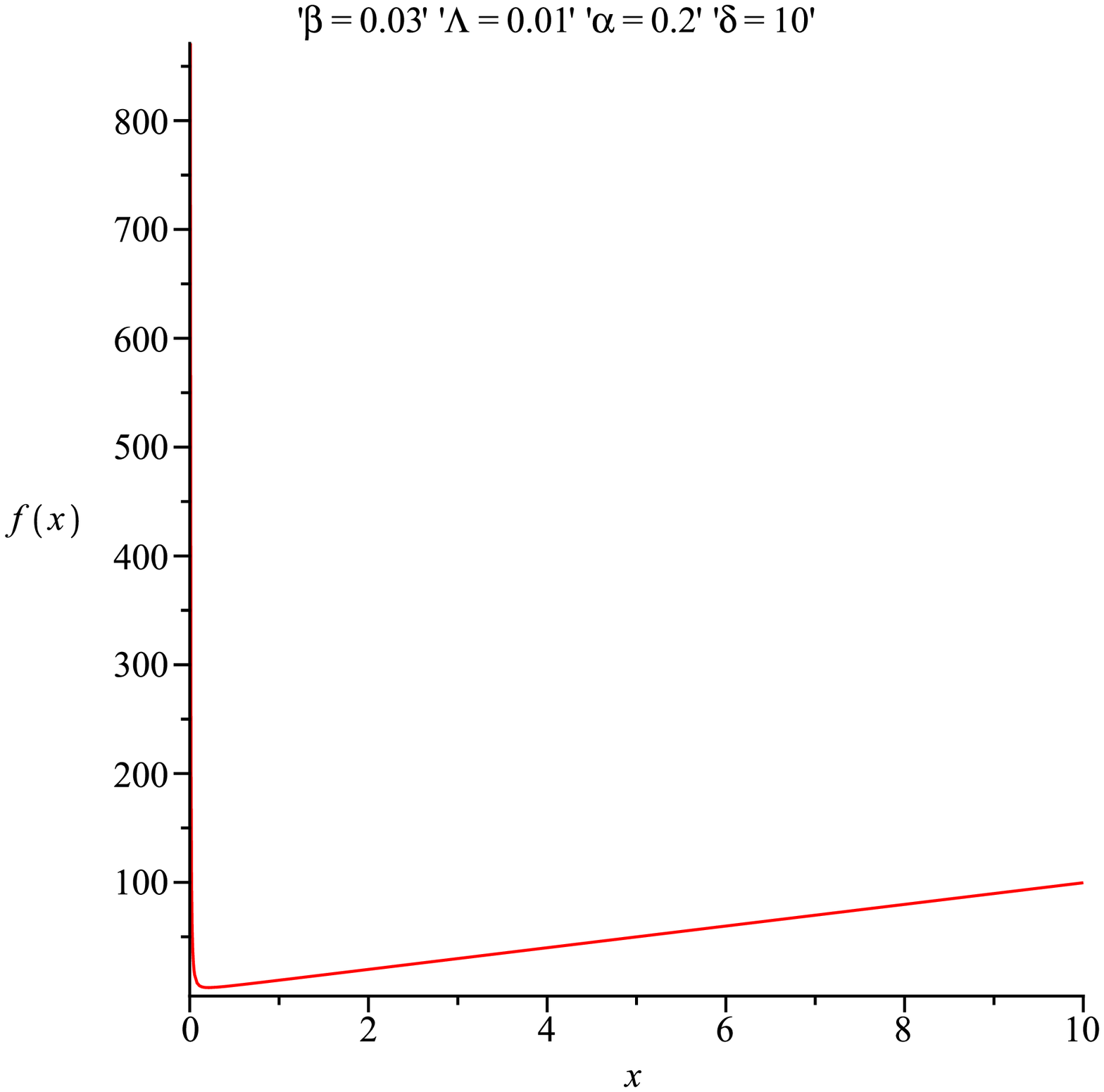}
\caption{Variation of f(x) with respect to x for positive values of $\beta$. (left panel) Anti de-Sitter spacetime (right panel ) de-Sitter spacetime.}
\end{figure*}
\begin{figure*}[tbp]
\centering
\includegraphics[width=6.5cm]{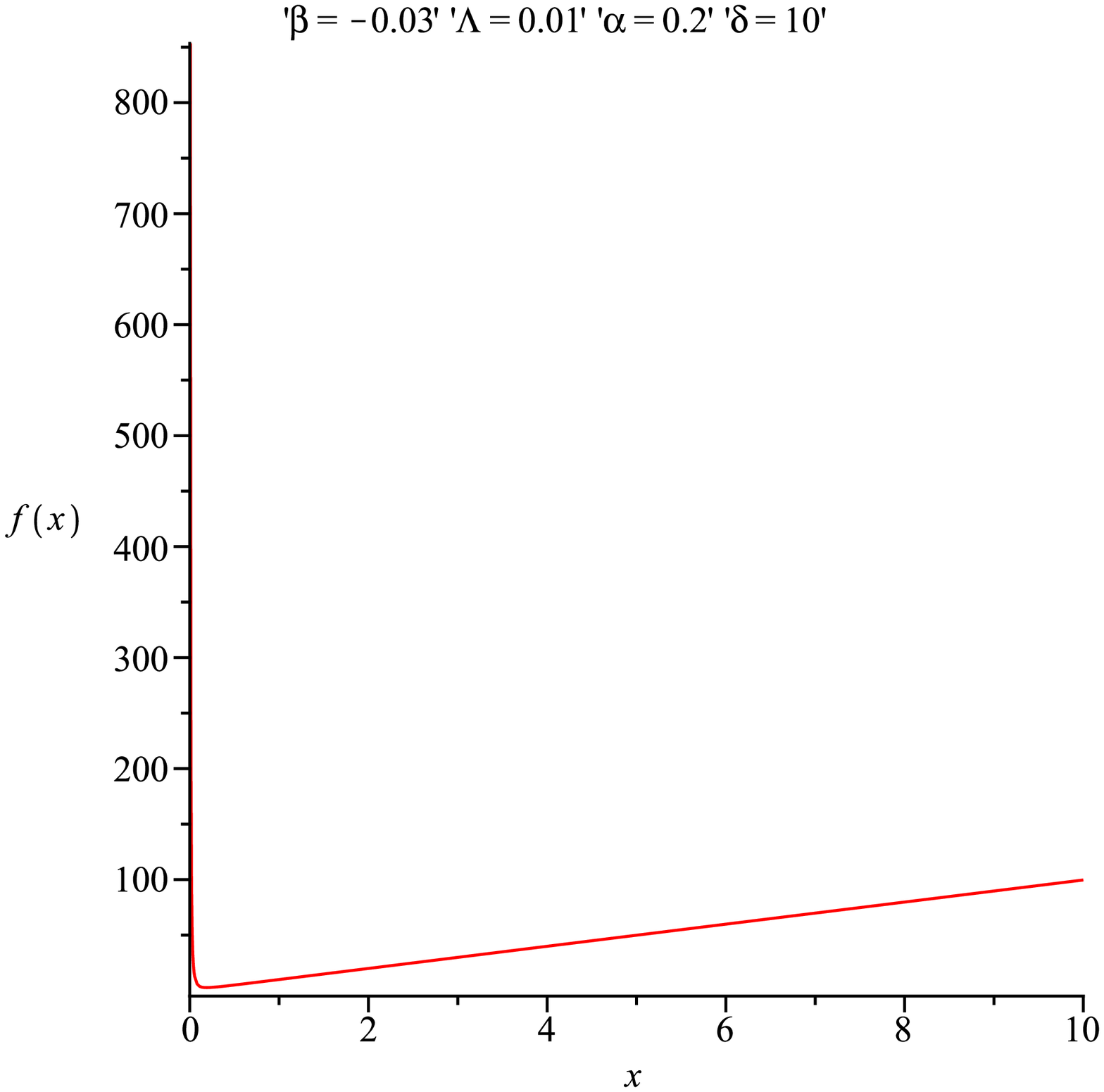}
\includegraphics[width=6.5cm]{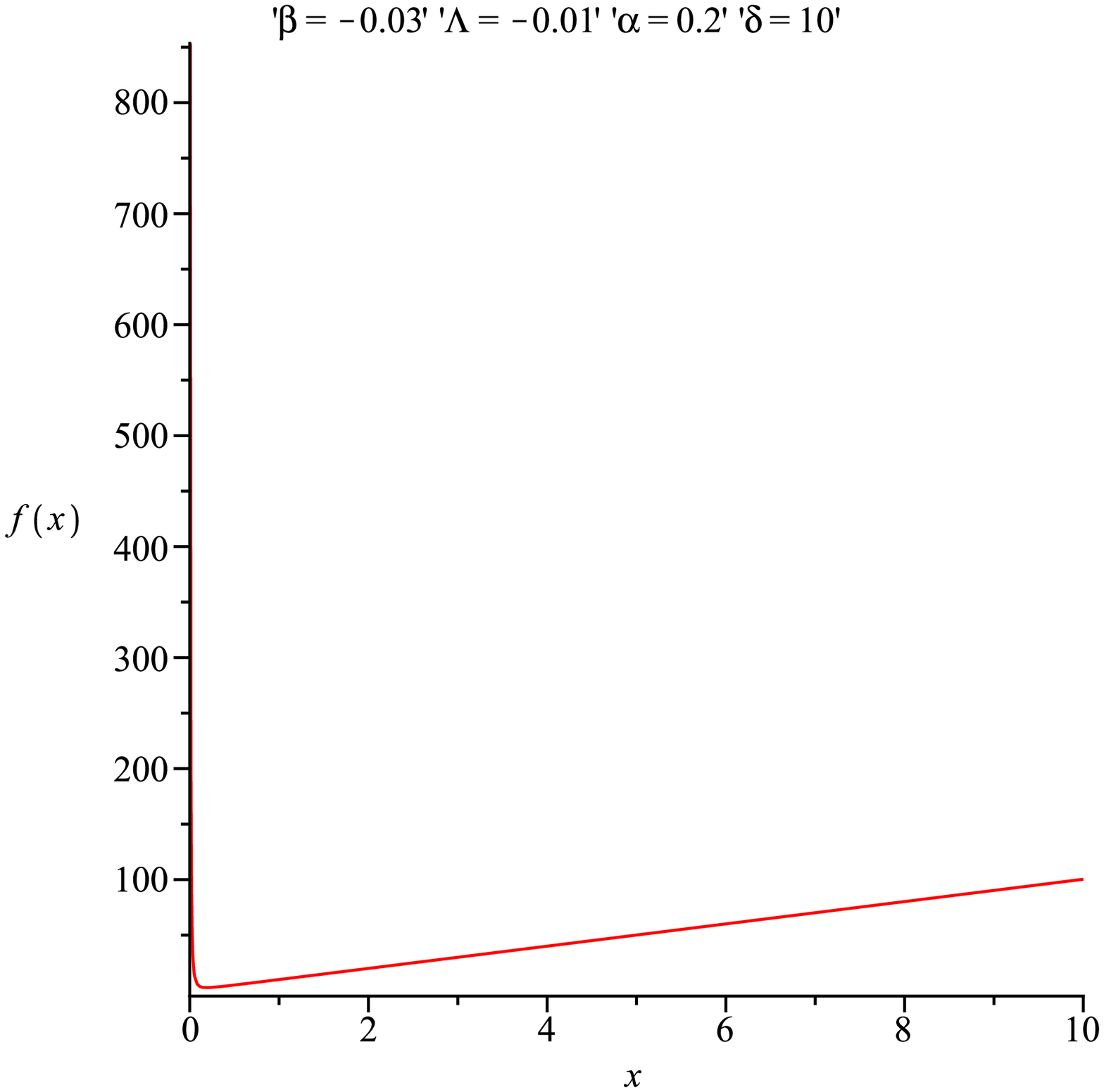}
\caption{Variation of f(x) with respect to x for negative values of $\beta$. (left panel) de-Sitter spacetime (right panel )  Anti  de-Sitter spacetime. }
\end{figure*}
\begin{figure*}[tbp]
\centering
\includegraphics[width=6.5cm]{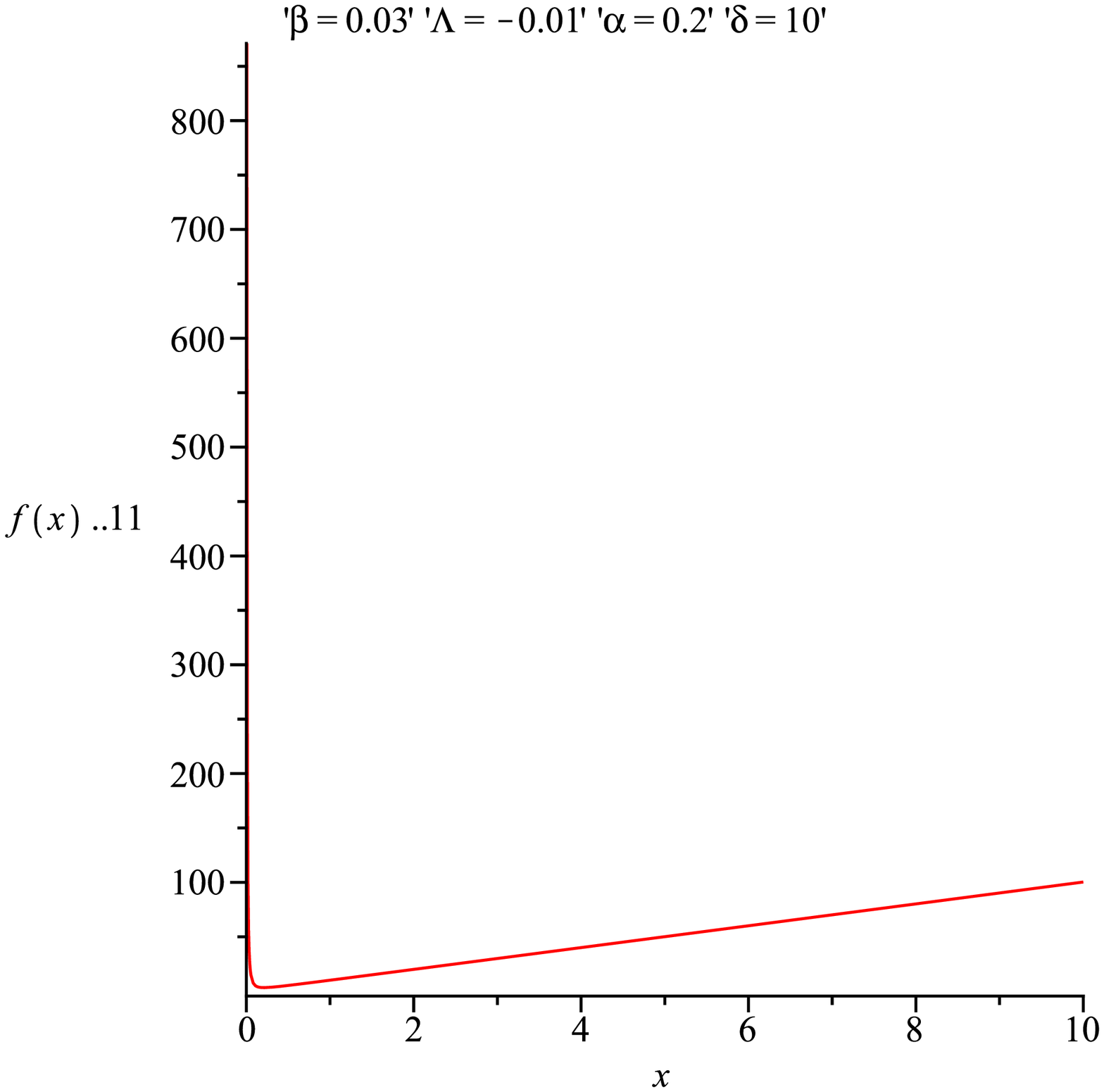}
\includegraphics[width=6.5cm]{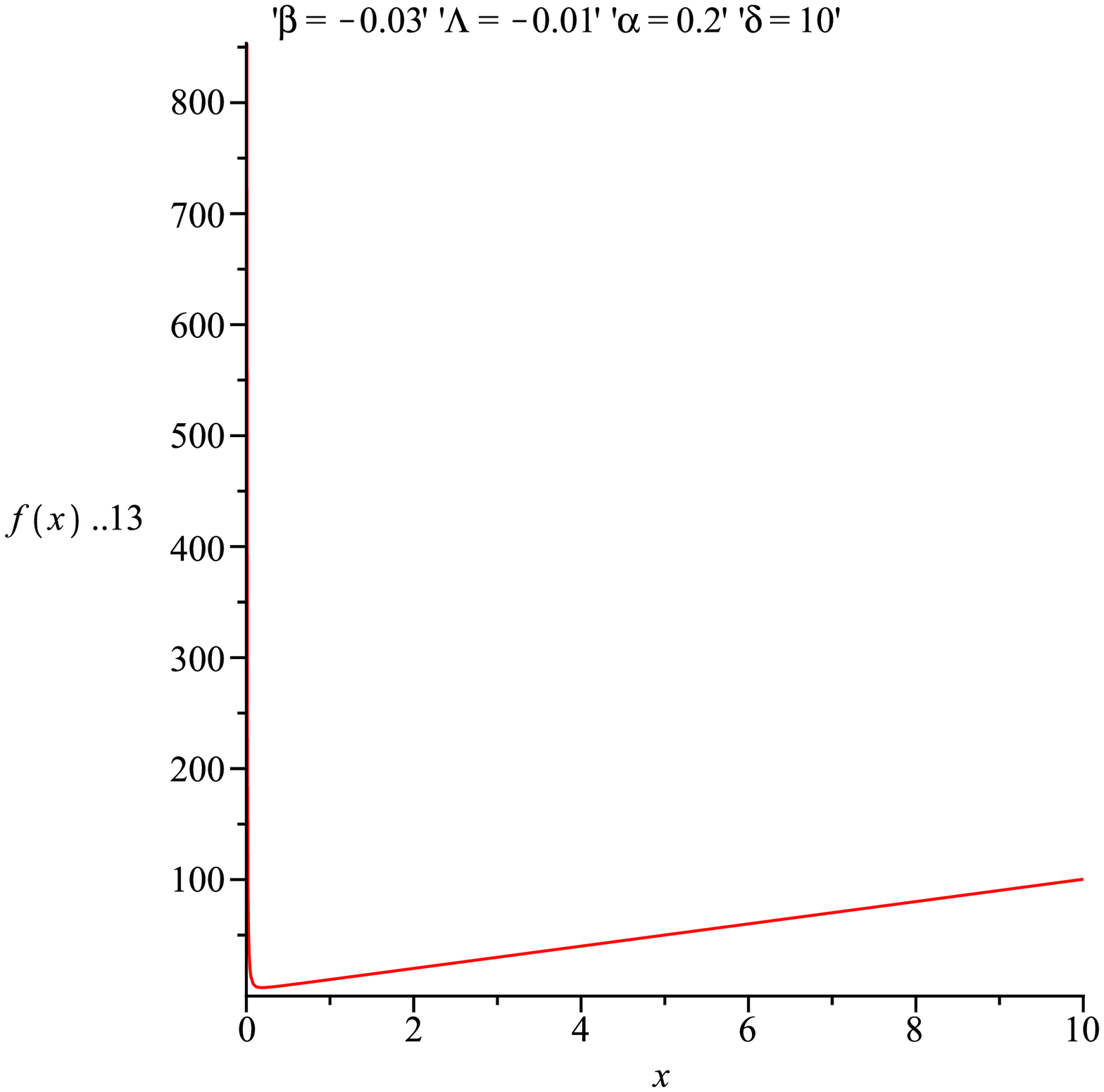}
\caption{Variation of f(x) with respect to x for Anti-de Sitter space. (left panel) Positive values of $\beta$ (right panel ) negative values of $\beta$. }
\end{figure*}

For $\delta=0$, the anisotropic fluid becomes a stiff fluid with the equation-of-state (EoS)
\begin{equation}
\kappa\,p=\kappa\,\rho=\frac{\gamma^2}{\frac{\Lambda}{3}\,x^2-(\frac{\alpha^2}{x^2}+\frac{2\,\beta}{x})}.
\label{final-final}
\end{equation}
Therefore the charged solution with a positive or negative cosmological constant represented by the metric (\ref{1}) possess naked singularities which are not clothed.

\section{The Energy-Momentum distributions}

Computing the conserved quantities, such as energy and momentum in curved space-times is still an unsolved problem. Following the Einstein's original pseudotensor for energy-momentum \cite{Ein,Ein1}, several expressions have been introduced in the literature, for instance, the Tolman \cite{Toll}, the Papapetrou \cite{Papa}, the Landau-Lifshitz's \cite{Lan}, the Bergman-Thomson \cite{Ber}, the M{\o}ller \cite{Moll,Moll2}, the Weinberg \cite{Wei}, and the Qadir-Sharif \cite{Qadir}. Rosen \cite{Rosen} and Cooperstock \cite{Cooper} calculated the energy and momentum distributions of a closed homogeneous isotropic universe described by Friedmann-Robertson-Walker (FRW) space-time using the Einstein complex and found that the total energy vanishes. Johri {\it et al} \cite{Johri} used the Landau-Lifshitz complex and found that the total energy of FRW spatially closed Universe vanishes. Many researchers considered different energy-momentum complexes in different space-times and obtained some encouraging results (see for example, \cite{Faiz8} and related references therein). In this article, we evaluate the energy-momentum distributions for the studied space-time (\ref{1}) using four different complexes.

\subsection{Landau-Lifshitz's Energy-Momentum Complex}

The energy and momentum densities in the sense of Landau-Lifshitz \cite{Lan} are given by
\begin{equation}
L^{\mu\rho}=\frac{1}{16\,\pi}\,S^{\mu\nu\rho\sigma}_{\,\,\,,\nu\sigma}
\label{aa1}
\end{equation}
where
\begin{equation}
S^{\mu\nu\rho\sigma}=-g\,(g^{\mu\rho}\,g^{\nu\sigma}-g^{\mu\sigma}\,g^{\nu\rho}),
\label{aa2}
\end{equation}
and has symmetries of the Riemann tensor $R_{\mu\nu\rho\sigma}$. Here $L^{00}$ represents the energy density and $L^{i0}$ represent the components of the momentum density, respectively. So, Landau-Lifshitz energy-momentum satisfies the local conservation laws as :
\begin{equation}
\frac{\partial L^{\mu\nu}}{\partial x^{\nu}}=0.
\label{aa3}
\end{equation}
The non-zero components of $S^{\mu\nu\rho\sigma}$ are
\begin{eqnarray}
S^{0101}&=&-x^4,\quad S^{0202}=-\frac{x^2\,e^{2\,\gamma\,t}}{f(x)},\quad S^{0303}=-\frac{x^2\,e^{-2\,\gamma\,t}}{f(x)},\nonumber\\
S^{1212}&=&x^2\,f(x)\,e^{2\,\gamma\,t},\quad S^{1313}=x^2\,f(x)\,e^{-2\,\gamma\,t}, \nonumber\\
S^{2323}&=&1.
\label{aa4}
\end{eqnarray}

Substituting (\ref{aa4}) in (\ref{aa1}), one will get the energy and momentum densities for the space-time (\ref{1}) given by
\begin{eqnarray}
L^{00}&=&\frac{1}{16\,\pi}\,S^{0\nu 0\sigma}_{\,\,\,,\nu\sigma}=\frac{1}{16\,\pi}\,S^{0101}_{\,\,,11}=-\frac{3\,x^2}{4\,\pi},\nonumber\\
L^{i0}&=&\frac{1}{16\,\pi}\,S^{i\nu 0\sigma}_{\,\,,\nu\sigma}=0,\quad i=1,2,3.
\label{aa5}
\end{eqnarray}
Note that the energy density components vanish at $x=0$ and remain same for all the above cases.

\subsection{Einstein Energy-Momentum Complex}

The energy-momentum complex as defined by Einstein is given by \cite{Ein,Ein1}
\begin{equation}
\Theta_{\mu}^{\,\,\,\,\nu}=\frac{1}{16\,\pi}\,H_{\mu\,\,\,,\rho}^{\,\,\,\,\nu\rho}
\label{bb1}
\end{equation}
where
\begin{equation}
H_{\mu}^{\,\,\,\,\nu\rho}=-H_{\mu}^{\,\,\,\,\rho\nu}=\frac{g_{\mu\tau}}{\sqrt{-g}}\,[-g\,(g^{\nu\tau}\,g^{\rho\sigma}-g^{\rho\tau}\,g^{\nu\sigma})]_{,\sigma}.
\label{bb2}
\end{equation}
The complex $\Theta_{\mu}^{\,\,\,\,\nu}$ satisfies the local conservation law as :
\begin{equation}
\frac{\partial \Theta_{\mu}^{\,\,\,\,\nu}}{\partial x^{\nu}}=0.
\label{bb3}
\end{equation}
Here $\Theta_{0}^{\,\,\,0}$ is the energy density, and $\Theta_{i}^{\,\,\,0}$ are the momentum density components.

The following components of $H_{\mu}^{\,\,\,\,\nu\rho}$ are needed
\begin{equation}
H_{0}^{\,\,\,01}=4\,x\,f(x),\quad H_{i}^{\,\,\,0\rho}=0.
\label{bb4}
\end{equation}
Using the above components in (\ref{bb1}), we get
\begin{equation}
\Theta_{\mu}^{\,\,\,\,0}=\frac{1}{16\,\pi}\,H_{\mu\,\,\,,\rho}^{\,\,\,\,0\rho}\nonumber\\
\end{equation}
Therefore the energy density component for the metric (\ref{1}) is
\begin{eqnarray}
\Theta_{0}^{\,\,\,0}&=&\frac{1}{16\,\pi}\,H_{0\,\,,\rho}^{\,\,\,\,0\rho}=\frac{1}{4\,\pi}\,[x\,f(x)]_{,x},\nonumber\\
&=&-\frac{1}{4\,\pi}\,(\frac{\alpha^2}{x^2}-2\,\delta\,x+x^2\,\Lambda).
\label{bb5-energy}
\end{eqnarray}
And the momentum components are
\begin{equation}
\Theta_{i}^{\,\,\,0}=\frac{1}{16\,\pi}\,H_{i\,\,,\rho}^{\,\,\,\,0\rho}=0.
\label{bb5}
\end{equation}

\subsection{Papapetrou Energy-Momentum Complex}

The symmetric energy-momentum complex of Papapetrou \cite{Papa} is given as
\begin{equation}
\Omega^{\mu\nu}=\frac{1}{16\,\pi}\,N^{\mu\nu\rho\sigma}_{\,\,\,,\rho\sigma}
\label{cc1}
\end{equation}
where
\begin{equation}
N^{\mu\nu\rho\sigma}=\sqrt{-g}\,(g^{\mu\nu}\,\eta^{\rho\sigma}-g^{\mu\rho}\,\eta^{\nu\sigma}+g^{\rho\sigma}\,\eta^{\mu\nu}-g^{\nu\sigma}\,\eta^{\mu\rho}),
\label{cc2}
\end{equation}
and $\eta^{\mu\nu}$ is the Minkowski space-time. The quantities $N^{\mu\nu\rho\sigma}$ are symmetric in its first two indices $\mu$ and $\nu$. So, Papapetrou energy-momentum satisfy the local conservation laws as :
\begin{equation}
\frac{\partial \Omega^{\mu\nu}}{\partial x^{\nu}}=0.
\label{cc3}
\end{equation}

The following components of $N^{\mu\nu\rho\sigma}$ are needed
\begin{eqnarray}
N^{0011}&=&-x^2\,[f(x)+\frac{1}{f(x)}],\nonumber\\
\quad N^{i0\rho\sigma}&=&-\sqrt{-g}\,(g^{i\rho}\,\eta^{0\sigma}+g^{0\sigma}\,\eta^{i\rho}),
\label{cc4}
\end{eqnarray}
where $i=1,2,3$. The energy and momentum densities are
\begin{eqnarray}
\Omega^{00}&=&\frac{1}{16\,\pi}\,N^{00\rho\sigma}_{\,\,\,,\rho\sigma}=\frac{1}{16\,\pi}\,N^{0011}_{\,\,\,,11} \nonumber\\
&=&-\frac{1}{16\,\pi}\,[x^2\,(f(x)+\frac{1}{f(x)})]_{,xx}\nonumber\\
\Omega^{i0}&=&\frac{1}{16\,\pi}\,N^{i0\rho\sigma}_{\,\,\,,\rho\sigma}=0.
\label{cc5}
\end{eqnarray}

The energy-density component for the space-time (\ref{1}) is given by
\begin{eqnarray}
\Omega^{00}&=&\frac{x}{8\,\pi}\,[-3\,\delta+2\,x\,\Lambda \nonumber\\
&&-\frac{27\,x\,(4\,\alpha^2+6\,\beta\,x+x^3\,\delta)^2}{(3\,\alpha^2+6\,\beta\,x+3\,x^3\,\delta-x^4\,\Lambda)^3}\nonumber\\
&&+\frac{9\,x\,(10\,\alpha^2+12\,\beta\,x+x^3\,\delta)}{(3\,\alpha^2+6\,\beta\,x+3\,x^3\,\delta-x^4\,\Lambda)^2}].
\label{cc6}
\end{eqnarray}
Notice that at $x=0$ the energy-density components vanish.

\subsection{M{\o}ller’s Energy-Momentum Complex}

The M{\o}ller's energy-momentum complex \cite{Moll,Moll2} is given by
\begin{equation}
M_{\mu}^{\,\,\,\,\nu}=\frac{1}{8\,\pi}\,\chi_{\mu\,\,,\rho}^{\,\,\,\,\nu\rho}
\label{dd1}
\end{equation}
where the superpotential $\chi^{\mu\rho}_{\nu}$ is
\begin{equation}
\chi_{\mu}^{\,\,\,\,\nu\rho}=-\chi_{\mu}^{\,\,\,\,\rho\nu}=\sqrt{-g}\,(g_{\mu\tau,\beta}-g_{\mu\beta,\tau})\,g^{\nu\beta}\,g^{\rho\tau}.
\label{dd2}
\end{equation}
The complex $M_{\mu}^{\,\,\,\,\nu}$ satisfies the local conservation laws :
\begin{equation}
\frac{\partial M_{\mu}^{\,\,\,\,\nu}}{\partial x^{\nu}}=0.
\label{dd3}
\end{equation}
Here $M_{0}^{\,\,\,\,0}$ and $M_{\alpha}^{\,\,\,\,0}$ are the energy and momentum density components, respectively.

The following components of $\chi_{\mu}^{\,\,\,\,\nu\sigma}$ are needed
\begin{eqnarray}
\chi_{0}^{\,\,\,\,0\rho}&=&-\sqrt{-g}\,g_{00,1}\,g^{00}\,g^{\rho 1},\quad \chi_{1}^{\,\,\,\,0\rho}=0,\nonumber\\
\chi_{2}^{\,\,\,\,0\rho}&=&\sqrt{-g}\,g_{22,0}\,g^{00}\,g^{\rho 2},\quad \chi_{3}^{\,\,\,\,0\rho}=\sqrt{-g}\,g_{33,0}\,g^{00}\,g^{\rho 3}.\nonumber\\
\label{dd4}
\end{eqnarray}
Using the above components in (\ref{dd1}), we obtain the energy and momentum densities
\begin{eqnarray}
M_{0}^{\,\,\,\,0}&=&\frac{1}{8\,\pi}\,\chi_{0\,,\rho}^{\,\,\,\,0\rho}=-\frac{x^2}{8\,\pi}\,f'(x),\nonumber\\
M_{i}^{\,\,\,\,0}&=&\frac{1}{8\,\pi}\,\chi_{i\,,\rho}^{\,\,\,\,0\rho}=0.
\label{dd5}
\end{eqnarray}
The energy-density component for the space-time (\ref{1}) is
\begin{equation}
M_{0}^{\,\,\,\,0}=\frac{1}{24\,\pi\,x}\,(6\,\alpha^2+6\,\beta\,x-3\,\delta\,x^3+2\,x^4\,\Lambda).
\end{equation}

\section{Discussions}

It is well known that exact solutions of the field equations have one of the mysterious features of the black hole which is called singularity. This singularity has been considered as one of the defects of the general relativity because explanation of singularity can not be made by the general relativity itself. Black holes became one of the most interesting object of study in gravitational physics already with the discovery of the Schwarzschild solution, the very first exact solution of Einstein's field equations. Although the singularity of the solution at the Schwarzschild radius turned out to be merely a coordinate singularity, rather than a singularity of the space-time, horizon is nevertheless a surface with very special and surprising properties. Oppenheimer and Snyder \cite{Opp} shown that during the collapse of spherically symmetric matter, a naked singularity and the event horizon will form. Penrose and Hawking \cite{Pen4,Hawking2} have formulated a set of results known as the {\it singularity theorems} which provide a powerful evidence that the formation of black holes is a generic feature of general relativity.

In the present work, a four dimensional non-static space-time with different stress-energy tensor is analyzed. Under various parameter conditions, we have seen that some solution (presented in {\it Case 1} to {\it Case 4}, sub-case (ii) of {\it Case 7}, and {\it Case 11}) possesses a naked singularity (NS) which is not covered by an event horizon. In these solution, various matter-energy sources, namely, isotropic perfect fluid, anisotropic fluid, stiff fluid etc. violates and/or satisfies the different energy conditions. By calculating the Kretschmann scalar (or the scalar curvature constructed from the Riemann tensor) we have seen that the singularity which is formed due to divergence of the scalar curvature or the matter energy-density, is naked singularity. Also by analysing the geodesic completeness condition in the space-time, we have found that the geodesic paths are incomplete for finite value of the affine parameter including $s=0$ and therefore the singularity is naked not clothed by an event horizon. In addition, some other solution possess a true curvature singularity which is covered by an event horizon and therefore these solution represents black hole models in the background of de-Sitter (dS) and anti-de Sitter (AdS) spaces depending on the signs of the cosmological constant. In {\it case} $1$, we have presented a stiff fluid solution with a naked singularity; in {\it case} $2$, an anisotroipc fluid solution with a naked singularity; in {\it case} $3$, a stiff fluid solution in the background of de-Sitter or anti-de Sitter spaces with a naked singularity; in {\it case} $4$, an anisotropic fluid solution with a naked singularity, and this solution reduces to conformally flat static solution as a special case. In {\it case} $5$, a non-static solution of stiff fluid which represents black hole model in the background of de-Sitter or anti-de Sitter (AdS) spaces, were presented. For static case, this black hole solution correspond to a de-Sitter or anti-de Sitter spaces depending on the signs of the cosmological constant. In {\it case} $6$, a static black hole solution with zero cosmological constant sourced by non-null electromagnetic field coupled with anisotropic fluid, were presented. In {\it case} $7$, a new algebraically general (Petrov type I) non-static solution of the Einstein's field equations representing a black hole model sourced by non-null electromagnetic field coupled with perfect fluid in the background of de-Sitter Universe, were presented. Under a very special case ($\gamma=0$), this non-static solution reduces to a Petrov type D static black holes model with only non-null electromagnetic field as the source and positive cosmological constant ($\Lambda>0$). For another special cases ($\gamma=0$, $\Lambda=0$), the non-static black holes solution reduces to a Petrov type D static model which possess a naked singularity (NS) with non-null electromagnetic field as the sources. In {\it case} $8$, zero cosmological constant non-static black holes solution sourced by a non-null electromagnetic field coupled with perfect fluid (stiff fluid), were presented. Under a special case ($\gamma=0$), this solution reduces to a static black hole model sourced by non-null electromagnetic field only. In {\it case } $9$, a non-static black holes solution with zero cosmological constant, and anisotroipc fluid as the matter-energy content, were presented. Under a special case ($\gamma=0$), this solution reduces to static black hole model with anisotroipc fluid. In {\it case} $10$, a non-static solution of the field equations representing black hole model with a positive or negative cosmological constant sourced by anisotropic fluid, were presented. Under a special case ($\gamma=0$), this solution reduces to static black hole models with anisotropic fluid in the background of de-Sitter or anti-de Sitter spaces. In {\it case} $11$, a non-zero cosmological constant solution of the field equations sourced by a non-null electromagnetic field coupled with anisotropic fluid possessing a naked singularity, were presented. By plotting graphs of the function $f(x)$ against $x$ for positive and negative cosmological constant, and choosing values of the different parameters, we have shown that the studied space-time (\ref{1}) possess naked singularities not covered by an event horizon.

Finally, we evaluate the energy-momentum distributions using the complexes of Landau-Lifshitz's, Einstein, Papapetrou, and M{\o}ller's prescriptions. We have observed that none of the prescriptions give the same energy-distributions for the given space-time, whereas the momentum-density components vanish. We also have observed that the stress-energy tensor $(T^{\mu}_{\nu}$) vanish at $x=0$ in some cases, and becomes infinite in some other cases including the electromagnetic field.

\section{ Acknowledgements:}

We would like to thank the kind referee(s) for the positive suggestions and valuable comments which have greatly improved the present text.

FR would like to thank the authorities of the Inter-University Centre for Astronomy and Astrophysics, Pune, India
for providing research facilities.  FR and SS are also grateful to DST-SERB (Grant No.:  EMR/2016/000193) and
UGC (Grant No.: 1162/(sc)(CSIR-UGC NET , DEC 2016)), Govt. of India, for financial support respectively.


\begin{thebibliography}{99}

\bibitem{Sakharov} A. D. Sakharov, Sov. Phys. JETP {\bf 22}, 241 (1966).
\bibitem{Gliner} E. B. Gliner, Sov. Phys. JETP {\bf 22}, 378 (1966).
\bibitem{Hawk} S. W. Hawking, Commun. Math. Phys. {\bf 25}, 152 (1972).
\bibitem{Friedmann} J. L. Friedman, K. Schleich and D. M. Witt, Phys. Rev. Lett. {\bf 71}, 1486 (1993).
\bibitem{Friedmann2} J. L. Friedman, K. Schleich and D. M. Witt, Phys. Rev. Lett. {\bf 75}, 1872 (1993).
\bibitem{Galloway} G. Galloway, K. Schleich, D. Witt and E. Woolgar, Phys. Rev. {\bf D 60}, 104039 (1999).
\bibitem{Cai} M. Cai and G.J. Galloway, Class. Quantum Grav. {\bf 18}, 2707 (2001).
\bibitem{Huang} C. Huang and C. Liang, Phys. Lett. {\bf A 201}, 27 (1995).
\bibitem{Lemos}  J. Lemos, Phys. Lett. {\bf B 353}, 46 (1995).
\bibitem{Cai2}  R. Cai and Y. Zhang, Phys. Rev. {\bf D 54}, 4891 (1996).
\bibitem{Mann}  R. B. Mann, Class. Quantum Grav. {\bf 14}, L109 (1997).
\bibitem{Brill} D. R. Brill, J. Louko and P. Peldan, Phys. Rev. {\bf D 56}, 3600 (1997).
\bibitem{Vanzo} L. Vanzo, Phys. Rev. {\bf D 56}, 6475 (1997).
\bibitem{Cai3} R. G. Cai, J. Ji and K. Soh, Phys. Rev. {\bf D 57}, 6547 (1998).
\bibitem{Klemm} D. Klemm, Class. Quantum Grav. {\bf 15}, 3195 (1998).
\bibitem{Klemm2} D. Klemm, V. Moretti and L. Vanzo, Phys. Rev. {\bf D 57}, 6127 (1998).
\bibitem{Smith} W. L. Smith and R. B. Mann, Phys. Rev. {\bf D 56}, 4942 (1997).
\bibitem{RG} R. G. Cai, L. Z. Qiao and Y. Z. Zhang, Mod. Phys. Lett. {\bf A 12}, 155 (1997).
\bibitem{JLemos} J. Lemos, Phys. Rev. {\bf D 57}, 4600 (1998).
\bibitem{Jos1} P. S. Joshi, {\it Global Aspects in Gravitation and Cosmology}, Clarendon Press, Oxford (1993).
\bibitem{Jos2} P. S. Joshi, {\it Singularities, Black Holes and Cosmic Censorship}, IUCAA publication, Pune, India (1997).
\bibitem{Tipler} F. J. Tipler, Phys. Lett. A {\bf 64}, 8 (1977).
\bibitem{Kro} A. Krolak, J. Math. Phys. {\bf 28}, 138 (1987).
\bibitem{Lei} G. Lemaitre, Ann. Soc. Sci. Bruxelles A {\bf 53}, 51 (1933).
\bibitem{Tol} R. C. Tolman, Proc. Natl. Acad. Sci. USA {\bf 20}, 169 (1934).
\bibitem{Bondi} H. Bondi, Mon. Not. Roy. Astron. Soc. {\bf 107}, 410 (1947).
\bibitem{pap} A. Papapetrou, in {\it A Random Walk in Relativity and Cosmology}, N. Dadhich {\it et al} (eds.), John Wiley \& Sons, New York (1985).
\bibitem{vai} P. C. Vaidya, Nature {\bf 171}, 260 (1953).
\bibitem{Christ} D. Christodoulou, Commun. Math. Phys. {\bf 93}, 171 (1984).
\bibitem{Des} S. S. Deshingkar, I. H. Dwivedi and P. S. Joshi, Phys. Rev. {\bf D 59}, 044018 (1999).
\bibitem{Gov} K. S. Govinder and M. Govender, Phys. Rev. {\bf D 68}, 024034 (2003).
\bibitem{Brave} S. Barve, T. P. Singh, Cenalo Vaz and L. Witten, Class. Quantum Grav. {\bf 16}, 1727 (1999).
\bibitem{Clarke} C. J. S. Clarke, Class. Quantum Grav. {\bf 10}, 1375 (1993).
\bibitem{Jos3} P. S. Joshi and I. H. Dwivedi, Phys. Rev. {\bf D 47}, 5357 (1993).
\bibitem{Jos4} P. S. Joshi and I. H. Dwivedi, Class. Quantum Grav. {\bf 16}, 41 (1999).
\bibitem{Glass} E. N. Glass and J. P. Krisch, Phys. Rev. {\bf D 57}, 5945 (1998).
\bibitem{Rocha} Jaime F. Villas da Rocha, arXiv: gr-qc/0105095.
\bibitem{Herra} L. Herrera, A. Di Prisco and J. Ospino, Eur. Phys. J. C. (2016) {\bf 76} : 603.
\bibitem{Kras} A. Krasinski, {\it Inhomogeneous Cosmological Models}, Cambridge University Press, Cambridge (1997).
\bibitem{Pen1} R. Penrose, Rivista del Nuovo Cimento {\bf 1}, 252 (1969).
\bibitem{Pen2} R. Penrose, {\it Singularities and time-asymetry}, in {\it General Relativity : An Einstein Centenary Survey}, S. W. Hawking {\it et al.} (eds.), Cambridge University Press, Cambridge (1979).
\bibitem{Pen3} R. Penrose, The Question of Cosmic Censorship, in {\it Black Holes and Relativistic Stars}, R. M. Wald (ed.), Chicago University Press, Chicago (1994).
\bibitem{Thor} K. S. Thorne, in {\it Magic without Magic : John Archibald Wheeler}, edited by J. R. Klauder, Freeman and Co., San Francisco (1972).
\bibitem{Hay} S. A. Hayward, Class. Quantum Grav. {\bf 17}, 1749 (2000).
\bibitem{Apo} T. A. Apostolatos and K. S. Thorne, Phys. Rev. {\bf D 46}, 2435 (1992).
\bibitem{Eche} F. Echeverria, Phys. Rev. {\bf D 47}, 2271 (1993).
\bibitem{Gutt} S. Guttia, T. P. Singh, P. A. Sundararaj and and C. Vaz, arXiv: gr-qc/0212089.
\bibitem{Nakao} K. Nakao and Y. Morisawa,  Class Quantum Grav. {\bf 21}, 2101 (2004).
\bibitem{Nakao1} K. Nakao and Y. Morisawa, Prog. Theo. Phys. {\bf 113}, 73 (2005).
\bibitem{Gonc} S. Goncalves and S. Jhingan, Int. J. Mod. Phys. {\bf D 11}, 1469 (2002).
\bibitem{Nolan} B. C. Nolan, Phys. Rev. {\bf D 65}, 104006 (2002).
\bibitem{Peri} P. R. C. T. Periera and A. Wang, Phys. Rev. {\bf D 62}, 124001 (2000).
\bibitem{Wang} A. Wang, Phys. Rev. {\bf D 68}, 064006 (2003).
\bibitem{Faiz5} F. Ahmed and F. Rahaman, Euro. Phys. J. A (2018) {\bf 54}: 52.
\bibitem{Faiz6} F. Ahmed and F. Rahaman, Adv. High Energy Phys. {\bf 2018}, 7839619 (2018).
\bibitem{Chi} T. Chiba, Prog. Theo. Phys. {\bf 95}, 321 (1996).
\bibitem{Piran} T. Piran, Phys. Rev. Lett. {\bf 41}, 1085 (1978).
\bibitem{Th} K. S. Thorne, Phys. Rev. {\bf 138}, B251 (1965).
\bibitem{Morgan} T. A. Morgan, Gen. Rel. Grav. {\bf 4}, 273 (1973).
\bibitem{Lete} P. S. Letelier and A. Wang, Phys. Rev. {\bf D 49}, 5105 (1994).
\bibitem{JMM} J. M. M. Senovilla and R. Vera, Class. Quantum Grav. {\bf 17}, 2843 (2000).
\bibitem{Bondi2} H. Bondi, Proc. R. Soc. Lond. {\bf A 427}, 259 (1990).
\bibitem{Mel} M. A. Melvin, Phys. Lett. 8, 65 (1964) ; Phys. Rev. 139, B225 (1965).
\bibitem{Wang2} J. C. N. de Araujo and Anzhong Wang, Gen. Rel. Grav. {\bf 32}, 1971 (2000).
\bibitem{Faiz} D. Sarma, F. Ahmed and M. Patgiri, Adv. High Energy Phys. {\bf 2016}, 2546186 (2016).
\bibitem{Faiz2} F. Ahmed, Adv. High Energy Phys. {\bf 2017}, 7943649 (2017).
\bibitem{Faiz3} F. Ahmed, Adv. High Energy Phys. {\bf 2017}, 3587018 (2017).
\bibitem{Faiz4} F. Ahmed, Prog. Theor. Exp. Phys. {\bf 2017}, 083E03 (2017).
\bibitem{Faiz7} F. Ahmed, Int. J. Geom. Meth. Mod. Phys. {\bf 15} (9), 1850153 (2018).
\bibitem{KS} K. S. Virbhadra, D. Narasimha, and S. M. Chitre, Astron. Astrophys. {\bf 337}, 1 (1998).
\bibitem{KS2} K. S. Virbhadra, G. F. R. Ellis, Phys. Rev. {\bf D 65}, 103004 (2002).
\bibitem{Chou} A. N. Chowdhury, M. Patil, D. Malafarina and P. Joshi, Phys. Rev. {\bf D 85}, 104031 (2012).
\bibitem{Zhou} S. Zhou, R. Zhang, J. Chen and Y. Wang, Int. J. Theor. Phys. {\bf 54}, 2905 (2015).
\bibitem{Yaz} G. N. Gyulchev and S. S. Yazadjiev, Phys. Rev. {\bf D 78}, 083004 (2008).
\bibitem{Sahu} S. Sahu, M. Patil, D. Narasimha, and P. S. Joshi, Phys. Rev. {\bf D 86}, 063010 (2012).
\bibitem{Mala} P. S. Joshi and D. Malafarina, Int. J. Mod. Phys. {\bf D 20}, 2641 (2011).
\bibitem{Mala2} P. S. Joshi and D. Malafarina and R. Narayan, Class. Quantum Grav. {\bf 28}, 235018 (2011).
\bibitem{Patil} M. Patil and P. S. Joshi, Phys. Rev. {\bf D 82}, 104049 (2010).
\bibitem{Patil2} M. Patil, P. S. Joshi, and D. Malafarina, Phys. Rev. {\bf D 83}, 064007 (2011).
\bibitem{Wer} M. Werner and A. Petters, Phys. Rev. {\bf D 76}, 064024 (2007).
\bibitem{Bambi} C. Bambi and N. Yoshida, Class. Quantum Grav. {\bf 27}, 205006 (2010).
\bibitem{Bambi2} C. Bambi and K. Freese, Phys. Rev. {\bf D 79}, 043002 (2009).
\bibitem{Hioki} K. Hioki and K. i. Maeda, Phys. Rev. {\bf D 80}, 024042 (2009).
\bibitem{Nara} P. S. Joshi, D. Malafarina, and R. Narayan, Class. Quantum Grav. {\bf 31}, 015002 (2014).
\bibitem{Hawking} S. W. Hawking and G. F. R. Ellis, {\it The Large Scale Structure of Space-Time}, Cambridge University Press, Cambridge (1973).
\bibitem{Wald} R. M. Wald, {\it General Relativity}, University of Chicago Press, Chicago (1994).
\bibitem{Steph} H. Stephani, D. Kramer, M. MacCallum, C. Hoenselaers and E. Herlt, {\it Exact Solutions to Einstein’s Field Equations}, Cambridge University Press, Cambridge (2005).
\bibitem{Wu} Y. Wu, M. F. A. da Silva, N. O. Santos and A. Wang, Phys. Rev. {\bf D 68}, 084012 (2003).
\bibitem{Taub} A. H. Taub, Ann. Math. {\bf 53}, 472 (1951).
\bibitem{Herr1} L. Herrera, G. Le Denmat, G. Marcilhacy and N. O. Santos, Int. Jour. Mod. Phys. {\bf D 14}, 657 (2005).
\bibitem{Herr2} L. Herrera, A. Di Prisco and J. Ospino, Gen. Rel. Grav. {\bf 44}, 2645 (2012).
\bibitem{PT} P. T. Chrusciel, http://homepage.univie.ac.at/pitor.chrusciel (BlackHolesViennaJuly2015.pdf, Revised July 2018), Erwin Schrodinger Institute and Faculty of Physics, University of Vienna.
\bibitem{Nielsen} A. B. Nielsen and J. H. Yoon, Class. Quantum Grav. {\bf 25}, 085010 (2008).
\bibitem{Valerio} V. Faraoni, {\it Galaxies} {\bf 1}, 114 (2013); doi:10.3390/galaxies1030114.
\bibitem{Roger-Penrose} R. Penrose, in {\it Batelle Rencountres}, C. M. DeWitt {\it et al} (eds.), Gordon and Breach, New York (1968).
\bibitem{Hay2} S. A. Hayward, Phys. Rev. {\bf D 49}, 6467 (1994).
\bibitem{Blau} M. Blau, http://www.blau.itp.unibe.ch/Lecturenotes.html (Lecture Notes on General Relativity, Last update August 8, 2016), Albert Einstein Center for Fundamental Physics, University of Bern, Switzerland.



\bibitem{Ein} A. Einstein, Sitzungsber. Preuss. Akad. Wiss. Berlin (Math. Phys.) {\bf 778} (1915).
\bibitem{Ein1} A. Trautman, {\it Gravitation: An Introduction to Current Research}, L. Witten (ed.), John Wiley \& Sons, New York (1962).
\bibitem{Toll} R. C. Tolman, {\it Relativity, Thermodynamics and Cosmology}, Oxford University Press, Oxford (1934).
\bibitem{Papa} A. Papapetrou, Proc. R. Irish Acad. {\bf A 52}, 11 (1948).
\bibitem{Lan} L. D. Landau and E. M. Lifshitz, {\it The Classical Theory of Fields}, Pergamon Press (1987).
\bibitem{Ber} P. G. Bergmann and R. Thompson, Phys. Rev. {\bf 89}, 400 (1953).
\bibitem{Moll} C. M{\o}ller, Ann. Phys. (N. Y.) {\bf 4}, 347 (1958).
\bibitem{Moll2} C. M{\o}ller, Ann. Phys. (N. Y.) {\bf 12}, 118 (1961).
\bibitem{Wei} S. Weinberg, {\it Gravitation and Cosmology: Principle and Applications of General Theory of Relativity}, John Wiley \& Sons, New York (1972).
\bibitem{Qadir} A. Qadir and M. Sharif, Phys. Lett. {\bf A 167}, 331 (1992).
\bibitem{Rosen} N. Rosen, Gen. Rel. Grav. {\bf 26}, 319 (1994).
\bibitem{Cooper} F. I. Cooperstock, Gen. Rel. Grav. {\bf 26}, 323 (1994).
\bibitem{Johri} V. B. Johri, D. Kalligas, G. P. Singh and C. W. F. Everitt, Gen. Rel. Grav. {\bf 27}, 313 (1995).
\bibitem{Faiz8} F. Ahmed, Euro. Phys. J. C (2018) {\bf 78}: 598.
\bibitem{Opp} J. R. Oppenheimer and H. Snyder, Phys. Rev. {\bf 56}, 455 (1939).
\bibitem{Pen4} R. Penrose, Phys. Rev. Lett. {\bf 14} 57 (1965).
\bibitem{Hawking2} S. W. Hawking and R. Penrose, Proc. R. Soc. {\bf A 314}, 529 (1970).


\end{thebibliography}
\end{document}